\documentclass[a4paper,fleqn]{cas-sc}

\usepackage{epsfig}
\usepackage{graphicx} 
\usepackage{bm}
\usepackage{soul}
\usepackage{amssymb}
\usepackage{subcaption}
\usepackage{caption}
\usepackage{adjustbox}
\usepackage{siunitx}
\usepackage{amsthm}
\usepackage{amsmath,amsfonts}
\usepackage{times}
\usepackage{caption}
\usepackage{subcaption}
\usepackage{lscape}
\usepackage{rotating}
\usepackage{epstopdf}
\usepackage[utf8x]{inputenc}
\usepackage[english]{babel}
\usepackage[version=4]{mhchem}
\usepackage{setspace}
\usepackage[12pt]{extsizes}
\usepackage{rotate}
\usepackage[title]{appendix}
\usepackage{hyperref}
\usepackage{subcaption} 
\usepackage[dvipsnames]{xcolor}
\usepackage{setspace}
\usepackage[normalem]{ulem}
\usepackage{booktabs} 
\usepackage{subcaption}

\usepackage{threeparttable}

\usepackage[authoryear]{natbib}
\usepackage[title]{appendix}

\def\tsc#1{\csdef{#1}{\textsc{\lowercase{#1}}\xspace}}
\tsc{WGM}
\tsc{QE}

\begin{document}
\let\WriteBookmarks\relax
\def\floatpagepagefraction{1}
\def\textpagefraction{.001}

\shorttitle{Dynamics of Large Impactors in a Magma Ocean}    

\shortauthors{L. Honarbakhsh et~al.}  

\title [mode = title]{Metal-Silicate Segregation During Planetary Accretion: Limited Iron Emulsification through intermediate Impacts}  



%

\author[1]{Leila Honarbakhsh}
\ead[url]{lhonarbakhsh@tulane.edu}

\credit{Conceptualization, Methodology, Numerical Modeling, Results Processing, Software development, Writing}

\affiliation[a]{organization={School of Science and Engineering, Tulane University},
                addressline={Blessey Hall}, 
                city={New Orleans},
                postcode={70118}, 
                state={Louisiana}, 
                country={United States}}

\affiliation[2]{organization={Department of Physics, University of Louisiana at Lafayette},
                addressline={Corner of Saint Mary Blvd. and Hebrard Blvd}, 
                city={Lafayette},
                postcode={70504}, 
                state={Louisiana},
                country={United States}}

\author[2]{Gabriele Morra}

\credit{Conceptualization, Methodology, Writing}
\author[3]{Peter Mora}

\credit{Numerical Model Development, Software Development, Writing}

\affiliation[3]{organization={School of Earth and Atmospheric Sciences, Queensland University of Technology, Brisbane, 4000, QLD},
                country={Australia}}

\author[1]{Colin RM. Jackson}
\credit{Conceptualization, Writing}


\cortext[1]{Corresponding author}

















\begin{abstract}
In this study, we investigate the post-collisional evolution of an impactor's iron core within a fully molten magma ocean using 2D numerical simulations with a novel Rothmann-Keller multiphase Lattice Boltzmann Method. The impactor's core diameter ranges from $152\;\text{km}$ to $552\;\text{km}$, with aspect ratios from $0.75$ to $5$. At Reynolds numbers up to $10^4$, our models reveal significant deformation and progressive fragmentation of the impactor into an iron cloud down to the smallest scale (several kilometers) that our method allows, which is then dispersed throughout the magma ocean by turbulent flow. We determined entrainment coefficients for a range of intermediate impactor sizes with various shapes, finding that larger impactors consistently exhibit higher entrainment coefficients. By extrapolating our mixing data, we anticipate incomplete iron-silicate mixing for kilometer-scale impactors in a real magma ocean, qualitatively consistent with previous predictions of partial equilibration. Additionally, our models highlight the mid- to lower magma ocean depths as critical zones for further iron fragment breakups and material transfer in the magma ocean.
\end{abstract}


\begin{highlights}
\item Entrainment coefficients increase with impactor size
\item Iron cloud geometrical dispersion remains limited for intermediate-sized impactors
\item Our models show that fragmentation continues for a longer duration after initial impact
\end{highlights}


\maketitle

\section{Introduction}
\label{sec:intro}

During planetary core formation, large differentiated impactors collided with the proto-Earth, generating magma oceans (MOs) whose depths and geometries varied based on the impactor’s size, velocity, and angle \citep{dale2023improved}. Researchers have extensively investigated the dynamics of the impactor core as it descended through these MOs using theoretical, numerical, and experimental methods \citep{rubie2003mechanisms, dahl2010turbulent, ichikawa2010direct, deguen2011experiments, samuel2012re, deguen2014turbulent, kendall2016differentiated, maas2021fate, maller2024condition, rohlen2025impactor}. The primary objective of these studies was to examine the metal-silicate segregation dynamics and to clarify the degree of metal-silicate mixing and equilibration, thereby refining the timing of accretion and enhancing the interpretation of geochemical signatures. 

Dominant turbulence within the MO plays a crucial role in segregation dynamics by strongly deforming the impactor's iron core and fragmenting it into a cloud of tiny droplets that disperse throughout the magmatic flow, seeding iron-silicate mixing. In effect, turbulent entrainment draws in the surrounding silicate, causing the iron cloud to expand as it descends to greater depths \citep{turner1986turbulent, deguen2014turbulent, kendall2016differentiated, landeau2021metal}. A larger amount of entrained MO facilitates more extensive chemical reactions, thereby exerting a stronger influence on the ultimate chemical composition of the mantle and core. The volume of the MO entrained by the iron cloud can be quantified by assuming a relationship between the entrainment rate and the inward entrainment velocity \citep{morton1956turbulent, turner1986turbulent}. According to the entrainment hypothesis, the entrained velocity across a turbulent flow is proportional to the flow's characteristic velocity \citep{morton1956turbulent, cenedese2024entrainment}, where the proportionality constant is the entrainment coefficient.

Dynamical conditions and the setup configuration are important factors that affect the entrainment coefficient. Various numerical and experimental investigations have estimated the entrainment coefficient under different dynamical conditions. Using laboratory thermals, \cite{turner1986turbulent} found that the entrainment coefficient for buoyant clouds and thermals ranges from $0.2$ to $0.25$. Similarly, \citet{deguen2014turbulent} reported entrainment coefficients of $0.2$-$0.3$ for interactions between immiscible fluids analogous to the iron-silicate system in a MO. Other studies have examined the effect of injector geometry on the entrainment rate and coefficient. For instance, numerical simulations by \cite{kandakure2005hydrodynamic} demonstrated that increasing the nozzle diameter enhances the air entrainment rate, a result that later was experimentally confirmed by \cite{balamurugan2006hydrodynamic}. Moreover, \cite{morrison2023controls} investigated the influence of the buoyant thermal aspect ratio on entrainment coefficient using direct numerical simulations at $\text{Re}=6300$, revealing its decrease by a factor of three as the aspect ratio increases by a factor of $4$. 

In this study, we simulate the interaction between two immiscible fluids, namely, the impactor's iron core and silicate in the MO, and similarly determine the effect of impactor geometry on the entrainment coefficient by testing various impactor diameters and aspect ratios.

Iron-silicate mixing and equilibrium efficiency in the MO depend on several factors such as the size of the impactor's core, depth of the MO, the iron droplet size, the amount of silicate in contact with iron, and the iron-silicate contact time \citep{rubie2003mechanisms, ichikawa2010direct, dahl2010turbulent, deguen2011experiments, deguen2014turbulent, rohlen2025impactor}. Some studies propose that small impactor cores, with diameters less than or equal to ten times the MO depth, fragment into centimeter-sized iron droplets within deep MOs, thereby facilitating efficient mixing and chemical equilibration \citep{deguen2014turbulent, landeau2014experiments}. In another study, \citet{samuel2012re} suggests that large diapirs smaller than the MO thickness may only reach chemical equilibrium over distances comparable to their initial size. According to \citet{dahl2010turbulent}, impactor cores exceeding $10\;\text{km}$ in diameter that collided with the proto-Earth during giant impacts may not fully disperse and equilibrate within the MO. In another study, \citet{kendall2016differentiated} modeled impacts of differentiated impactors with diameters of $100\;\text{km}$ using hydrocode simulations. They showed that kilometer-scale impactor fragments can achieve chemical equilibrium in the deeper regions of the MO through turbulent entrainment. Thus, fragment size and MO depth are key parameters in assessing chemical equilibration, together with the contact time between the iron and silicate phases.

Impact dynamics is another critical factor on the impactor's core fragmentation and iron-silicate mixing. \cite{rohlen2025impactor} investigated the influence of impactor size and velocity on its fragmentation within the MO using hydrocode models. Their findings revealed that impactor fragmentation is strongly dependent on MO depth, with shallower MOs exhibiting a lower degree of fragmentation and chemical equilibration. Similarly, \cite{maller2024condition} experimentally examined the effects of impactor velocity and size on fragmentation in the MO, identifying the Froude number as a key factor controlling the fragmentation process. By experimentally investigating the effects of impact dynamics, including impactor speed, size, and density, as well as crater formation and jet collapse inertia, \cite{landeau2021metal} found that undifferentiated impactors with diameters $<100\;\text{km}$ exhibit greater mixing with full chemical equilibration, whereas giant impacts show partial equilibration. Thus, the dynamics of the impactor core-MO interaction are central to understanding the chemical equilibration history of the early Earth.

In this study, we pioneer the modeling of immiscible multiphase flow to explicitly simulate the interaction of an iron impactor core with a MO flow, independent of pre-existing thermal convection. We utilize a two-phase Lattice Boltzmann Method (LBM) \citep{mora2021optimal,mora2024progress} designed for immiscible fluids, which enables the self-consistent modeling of iron-silicate differentiation. By assigning discrete physical properties, such as density and viscosity, to each phase and utilizing a sophisticated 'recoloring' algorithm \citep{mora2021optimal}, this numerical framework accurately resolves topological transitions from a coherent impactor to fragmented iron bodies. By maintaining sharp phase boundaries, this approach allows us to closely monitor the impactor’s fragmentation dynamics and the resulting extent of iron-silicate emulsification within the MO. 

A key challenge in modeling the interaction between the descending impactor core and the dynamics of the MO is accounting for the extremely turbulent nature of the MO. This turbulence is characterized by a computationally prohibitive high Reynolds number ($\approx 10^{15}$). By focusing on the Reynolds number ranges up to $10^4$ in our study, we ensure a balance between computational feasibility and fidelity to the underlying physical processes, such as impactor fragmentation and interaction with MO turbulence.

Quantifying iron-silicate mixing is both challenging and essential for accurately depicting the chemical and compositional state of our planet. Numerous studies have examined various mixing indices across different applications, with the standard deviation of the sample concentration relative to its mean serving as the most common indicator of the mixing process \citep{levy2001handbook}. \citet{poux1991powder} and \citet{wang1977stochastic} explored and tabulated various solid mixing indices, emphasizing the interrelation among these indices and highlighting the need to consider sample size in some cases. Similarly, \citet{doroodchi2013liquid} and \citet{hashmi2014quantification} investigated the mixing behavior in microfluidic systems, examining several variance-based mixing indices. In another study, \citet{kazemzadeh2021analysis} employed a variance-based mixing index to evaluate the performance of a high-speed Scott impeller in mixing two immiscible liquids. Likewise, \citet{duan2019experimental} evaluated the mixing performance of an immiscible liquid–liquid cyclone reactor using variance-based mixing indices, whereas \citet{zhu2024study} defined mixing entropy and mixing time based on information entropy to measure the mixing degree of ionic liquid alkylation. The defined mixing metrics vary from $0$ to $1$, representing no mixing to perfect mixing, respectively. 
In this work, we use mixing entropy analyses to investigate the extent of iron-silicate mixing induced by the interaction of the impactor’s iron core with the MO across a range of diameters and aspect ratios.

The structure of the paper is as follows: Section \ref{sec:method} outlines the simulation methodology, Section \ref{sec:result} presents the simulation results, and Sections \ref{sec:discussion} and \ref{sec:conclusion} provide the discussion and conclusion, respectively.



\section{Method}
\label{sec:method}
\subsection{Lattice Boltzmann method}
\noindent
The lattice Boltzmann method (LBM) has become increasingly prominent in computational fluid dynamics owing to its capability to model nonlinear rheologies, along with its straightforward parallelization, excellent scalability to large core counts, and efficient treatment of interfacial dynamics and complex boundaries \citep{he1999lattice, chiappini2010improved, liang2016lattice, mora2021optimal, mora2023hpc, Mora2023-exascale, mora2025geo}.

The iron-silicate segregation dynamics relevant to planetary core formation are primarily governed by Rayleigh-Taylor instability, which drives the descent of iron-rich diapirs or droplets through the silicate layer of the MO. Secondary Plateau-Rayleigh instabilities can further fragment iron streams and filaments into droplets as segregation progresses \citep{ fleck2018iron}. To this end, the Rothman–Keller multiphase lattice Boltzmann method (RKLBM) \citep{mora2021optimal} provides a robust computational framework for investigating these processes \citep{latva2005diffusion, saito2017lattice, haghani2024color}. In this study, we employ the RKLBM framework to simulate iron-silicate segregation under conditions relevant to early planetary differentiation. This framework can be executed by excluding the thermal convection from RKTLBM (RK-Thermal LBM), which can model both thermal convection and multiphase flow simultaneously. However, its parameter interrelations and scaling still benefit from the use of thermal parameters (discussed later). The methodology formulation is outlined below.

In the LBM terminology, the $DnQm$ discrete lattice refers to a system with $n$ dimensions and $m$ discrete velocity vectors (the $D2Q9$ is used in this study). The velocity vectors, $\bf c_{\alpha}$, and weights, $\omega_{\alpha}$, at each Cartesian lattice node and direction, $\alpha$, are defined as
\begin{equation}
{\bf c}_\alpha \ = \
\begin{array}{l}
[( 0, 0),\\
( 1, 0),( 0, 1),(-1, 0),( 0,-1),\\
( 1, 1),(-1, 1),(-1,-1),( 1,-1)] \Delta x / \Delta t
\end{array}
 \ \ \
\label{eq:velocity2D} \ \ \ ,
\end{equation}
\noindent 
and
\noindent
\begin{equation}
{w}_\alpha \ = \
\begin{array}{l}
[4/9,\\
1/9,1/9,1/9,1/9,\\
1/36,1/36,1/36,1/36]
\end{array}
 \ \ \
\label{eq:velocity2D} \ \ \ ,
\end{equation}
\noindent
where $\Delta \text{x}$=$1$ denotes the lattice spacing, and $\Delta \text{t}$=$1$ is the lattice timestep. The corresponding velocity and weight parameters are used throughout the framework as described below. 

RKLBM is implemented through three fundamental steps: (i) streaming, (ii) collision, and (iii) recoloring. In the streaming step, the particle density distribution functions, $f_{\alpha}^{k}$ ($k=1,2$, and $k$ refers to the iron and silicate phases), propagate along the lattice directions in each lattice node, and the following computational steps are implemented at each timestep to determine the momentum transfer. 

The post-streaming distribution function is given by
\noindent
\begin{equation}
f_{\alpha}^{k}(\mathbf{x},t) = f_{\alpha}^{k}(\mathbf{x}-\mathbf{c_{\alpha}} {\Delta t,t-\Delta t})
\end{equation}
\noindent
from which the macroscopic properties, such as fluid density and velocity, are obtained using the following expressions:
\begin{equation}
\rho_{k} = \sum_{\alpha} f_{\alpha}^{k} \ \ \ ,
\end{equation}
\begin{equation}
\rho = \sum_{k} \rho^{k} \ \ \ ,
\end{equation}
\begin{equation}
\mathbf{u} = \frac{1}{\rho}\sum_{k}\sum_{\alpha} f_{\alpha}^{k}\mathbf{c_{\alpha}} \ \ \ ,
\end{equation}
\noindent

In the collision step, the standard LBM collision term relaxes the distributions toward Boltzmann equilibrium, while an additional term captures cohesive forces between the fluids. 
The collision step is obtained by adding the collision terms to the streaming terms:
\begin{equation}
f_{\alpha}^{k*}(\mathbf{x},t)= f_{\alpha}^{k}(\mathbf{x},t)+(\Delta f_{\alpha}^{k})^{1}+(\Delta f_{\alpha}^{k})^{2} \ \ \ ,
\end{equation}
\noindent

where $^*$ superscript represents the post-collision distributions, and $(\Delta f_{\alpha}^{k})^1$ and $(\Delta f_{\alpha}^{k})^2$ correspond to the first and second collision terms, respectively. These terms and their constituent components are defined in the following:

\noindent
\begin{equation}
(\Delta f_{\alpha}^{k})^1 = \frac{1}{\tau}[f_{\alpha}^{k,eq}(x,t)-f_{\alpha}^{k}(x,t)]
\ \ \ ,
\end{equation}
\noindent
\begin{equation}
(\Delta f_{\alpha}^{k})^2 = A|\mathbf{F}|\left\{\omega_{\alpha}[\cos(\lambda_{\alpha})|\mathbf{c_{\alpha}}|]^2 - B_\alpha \right\}
\label{eq:second_collision}
\end{equation}
\noindent
Here, $\mathbf{F(\mathbf{x},t)}= \Sigma_{\alpha}\mathbf{c}_\alpha[\rho_{1}(\mathbf{x}+\mathbf{c}_{\alpha}\Delta T,t)-\rho_{2}(\mathbf{x}+\mathbf{c}_{\alpha}\Delta T,t)]$ denotes the color gradient \citep{latva2005static}, $\lambda_{\alpha}$ is the angle between $\mathbf{F}$ and $\mathbf{c}_{\alpha}$, and defined as $\cos(\lambda_{\alpha})=\frac{\mathbf{c}_{\alpha}.\mathbf{F}}{|\mathbf{c}_{\alpha}||\mathbf{F}|}$. $B_{0}=-4/27$, $B_{\alpha}=2/27$ for $\alpha=1,2,3,4$, and $B_{\alpha}=5/108$ for $\alpha=5,6,7,8$.

$\tau$ is the relaxation time, representing the characteristic duration over which the density distribution function relaxes toward its respective equilibrium state. For the momentum distribution functions, the relaxation time controls the rate of momentum diffusion and is related to the kinematic viscosity as:

\begin{equation}
\tau_{f}^{k}= 3\nu_{f}^{k}+0.5\ \ \,
\end{equation}
where $\nu_f^{k}$ denotes the kinematic viscosity of each phase. To manage abrupt variations in the relaxation time, $\tau$, at the phase interface, different non-linear relations of $\psi(x)= \frac{\rho_{1}(x)-\rho_{2}(x)}{\rho_{1}(x)+\rho_{2}(x)}$ are used to smooth the relaxation time between the two phases \citep{grunau1993lattice, mora2021optimal}.

$f_{\alpha}^{k,eq}(x,t)$ denotes the density equilibrium distribution function, which the distribution functions relax towards, and defined as follows:

\begin{equation}
f_{\alpha}^{k,eq}(x)= \rho_{k}(P_{\alpha}^{k}+\omega_{\alpha}[3(\mathbf{c}_{\alpha}.\textbf{u})+\frac{9}{2}(\mathbf{c}_{\alpha}.\textbf{u})^{2}-\frac{3}{2}\textbf{u}^2])\ \ \ ,
\end{equation}
\noindent
where 
\noindent
\begin{equation}
P_{\alpha}^{k} = 
    \begin{cases}
        \alpha_{k}, & \alpha= 0 \\
        \frac{1-\alpha_{k}}{5}, & \alpha= 1,2,3,4 \\
        \frac{1-\alpha_{k}}{20}, & \alpha= 5,6,7,8 \\
    \end{cases}
\end{equation}
\noindent
and $\alpha_{1}= \frac{\rho1}{\rho1+\rho2}$ and $\alpha_{2}= \frac{\rho2}{\rho1+\rho2}$ \citep{grunau1993lattice, reis2007lattice}. $P_{\alpha}^k$ denotes the rest and moving distributions in each phase when $\bf{u}$=$0$.

After applying the streaming and collision steps to $f_{\alpha}^{k}$, which carries mass density of fluid $k$, one obtains the incompressible Navier-Stokes equations \citep{chen1998lattice, succi2001lattice, dellar2013interpretation}.
The last step of the multiphase RKLBM to achieve the separation of the two immiscible fluids is called recoloring and is obtained as follows \citep{rothman1988immiscible}:
\begin{equation}
f_{\alpha}^{k**}(\mathbf{x},t)  = \frac{\rho_{1}}{\rho}f_{\alpha}^{*}+\zeta\frac{\rho_{1}\rho_{2}}{\rho^2}f_{\alpha}^{eq}(\rho,\mathbf{u}=0)cos(\lambda_{\alpha})\pm F_{k\alpha}^b \ \ \ ,
\label{eq:beta}
\end{equation}
\noindent
where $^{**}$ superscript on the left-hand side refers to the post-recoloring distribution, $f^*_\alpha = \sum_k f_\alpha^{k*}$, and $\zeta$ $\in(0,1]$ is a parameter that can be adjusted to control the interfacial thickness ($0.75$ in this study). $F_{k\alpha}^b$ accounts for the buoyancy due to both temperature and compositional density differences. In other words, $F_{k\alpha}^b$ is a force term in the LBM in direction $\alpha$ which applies an appropriate buoyancy force that captures the force on each fluid due to temperature differences (Boussinesq term) as well as compositional density differences (\textit{i.e.} $\Delta \rho$) where $\Delta \rho$ = $\rho - \rho_{2}$ is the density difference between the total fluid and pure silicate.

\subsection{Numerical Setup}
\label{sec:model setup}
\begin{figure*}[h] 
\centering
\includegraphics[trim=10 10 10 10,clip,width=0.5\linewidth]{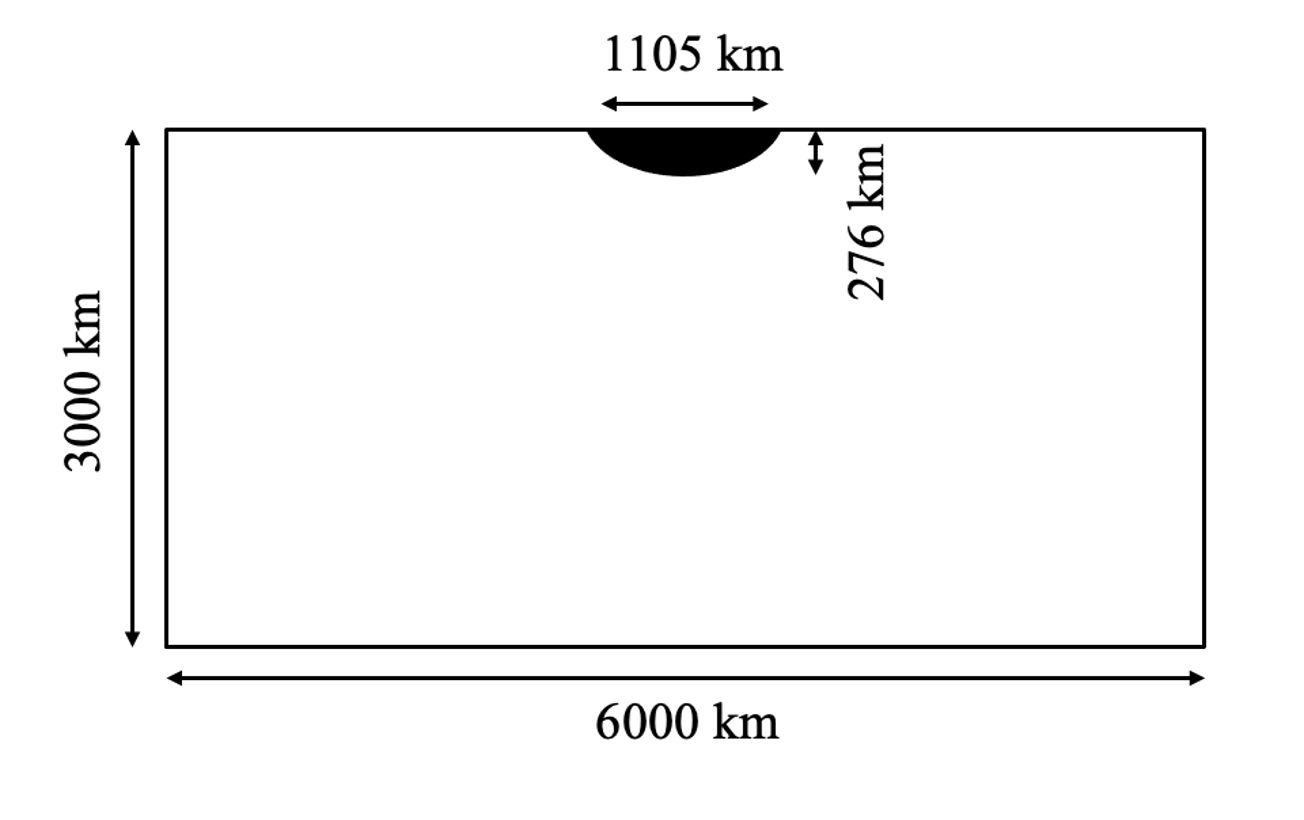}
\caption{Schematic representation of our simulation model setup. The impactor is initially positioned at the top of the MO.}
\label{fig:model_setup}
\end{figure*}

\noindent
In this study, we investigate the post-impact evolution of an impactor's differentiated iron core (hereafter referred to as the impactor) in a turbulent MO while varying the impactor size and aspect ratio. We analyzed the temporal evolution of intermediate impactors with diameters: $152\;\text{km}$, $252\;\text{km}$, $352\;\text{km}$, $452\;\text{km}$, and $552\;\text{km}$ in the MO. The aspect ratio, defined as the ratio of the impactor's width to its height, ranges from $0.75$ to $5$, with increments of $0.25$.
One example of our model setup is shown in Figure \ref{fig:model_setup} for the impactor with a diameter of $552\;\text{km}$ and aspect ratio of $5$. In our models, we maintain a fixed aspect ratio of $2$ for the numerical domain. For a given aspect ratio of $5$, the impactor is modeled as a semi-ellipse with horizontal and vertical axes of $1105\;\text{km}$ and $276\;\text{km}$, respectively, corresponding to a circular impactor with a diameter of $552\;\text{km}$.

Previous planetesimal impact studies \citep{kendall2016differentiated} indicate that an impactor with a diameter greater than $10\;\text{km}$ remains intact after the impact and eventually comes to rest before interacting with the turbulent MO. Accordingly, we positioned the impactor near the surface at zero initial velocity in our simulations to investigate its emulsification and interactions within the MO.
In our models, we implemented lateral periodic boundary conditions along with free-slip conditions at the top and bottom. Periodic lateral boundaries eliminate artificial edge effects and prevent the introduction of shear stresses that are not present in the actual MO. Meanwhile, the free-slip conditions at the top and bottom allow the impactor, initially placed at the top of the domain, to interact naturally with the underlying MO layer, ensuring a realistic representation of its engagement with deeper regions without the interference of artificial friction at the bottom boundary.

Table \ref{tab:table1} outlines the MO and simulation parameters used in this study.
To characterize the dynamic behavior of our models, which mimic the interaction between an impactor and the MO, we introduce several dimensionless parameters, such as Reynolds number ($\text{Re}$), Weber number ($\text{We}$), Bond number ($\text{Bo}$), and Froude number ($\text{Fr}$), along with density and viscosity ratios. These parameters, discussed in detail below, define the dynamic regimes of iron-silicate interactions within the MO.

The Reynolds number ($\text{Re}$), which represents the ratio of inertial to viscous forces, is defined as $\text{Re}=ul/\nu$ where $\nu$ is the kinematic viscosity, $l$ is the impactor's initial radius, and $u$ is the sinking velocity. In the real MO, $\text{Re}$ can reach $10^{15}$. This estimate assumes an impactor with a $100\;\text{km}$ radius, a post-impact velocity of several $\text{km/s}$ \citep{deguen2011experiments}, and the silicate kinematic viscosity between $10^{-7}-10^{-4}$ $\text{m}^2/\text{s}$ (table \ref{tab:table1}). In our simulations, the core has a diameter ranging from $208$ to $754$ pixels, equivalent to $152\;\text{km}$ and $552\;\text{km}$, the smallest and largest impactor diameters, respectively. Using a model (non-dimensional) viscosity of $\nu = 8.28\times10^{-4}$ and sinking velocity slightly above $10^{-2}$, the resulting $\text{Re}$ fall within the range $5896$--$61025$. For a fixed impactor diameter, the $\text{Re}$ values remain of the same order of magnitude, irrespective of the aspect ratio employed in the model.

The Weber number is defined as $\text{We}=\frac{\rho_{fe}u^{2}l}{\sigma}$, and the Bond number is given by $\text{Bo}=\frac{\Delta\rho gl^{2}}{\sigma}$, where $\rho_{fe}$ is the iron density, and $\Delta \rho$ and $\sigma$ are the density difference and surface tension between iron and silicate, respectively. In real planetary impacts involving bodies with radii of $100\;\text{km}$ or more, the MO naturally experiences $\text{We}$ and $\text{Bo}$ values on the order of $10^{14}$ or beyond \citep{deguen2011experiments}. In our models, $\sigma$ is set to $1.16\times 10^{-4}$ in lattice unit, resulting in $\text{We}$ and $\text{Bo}$ values in the range $567-894$ and $99-1299$, respectively, across various impactor sizes. Notably, the typical fragment radius in our simulations (determined as the most frequent radius in radii distribution over the entire simulation time) is $5$ pixels (approximately $7\;\text{km}$, discussed in Section \ref{sec:frag}), and $\sigma$ in lattice units for a single circular iron fragment of that size is determined from the 2D Young–Laplace relation $dP = \sigma / r$, where $dP$ is the pressure difference between the inside and outside of the fragment.  Despite the difference in scale between the $\text{We}$ and $\text{Bo}$ values in our models and those of real MO, the iron fragmentation regime observed in our simulations is qualitatively consistent with the reported fragmentation in laboratory experiments \citep{deguen2011experiments, deguen2014turbulent, landeau2014experiments, landeau2021metal}. This agreement indicates that our models capture the essential physics governing impactor segregation within the MO and are therefore capable of reproducing results consistent with experimental investigations.


We define the viscosity ratio as the ratio between the viscosity of liquid silicates and that of liquid iron. Typically, liquid iron has a viscosity in the range of $5.4\times10^{-3}$ to $2\times10^{-3}$ \text{Pa.s} \citep{assael2006reference}, while liquid silicates range from $10^{-2}$ to $10^{-1}$ \text{Pa.s} under conditions spanning atmospheric pressure to the core-mantle boundary \citep{bajgain2022insights}. This yields a relative viscosity for the MO generally between $10$ and $100$. However, for simplicity, our simulations set the viscosities of iron and silicate equal, even though in reality silicate's viscosity is about an order of magnitude greater than that of liquid metal. The scaling of our numerical experiments justifies this assumption, as explained below. Mapping our high-resolution grid to the full MO depth ($\approx3000\;\text{km}$) yields a pixel width of about $732$ meters. At this scale, each LBM fragment corresponds to an aggregate of many iron droplets within a predominantly silicate medium. Thus, the viscosity assigned to each computational cell is an average weighted mainly by the silicate fraction, making the effect of lower iron viscosity negligible.

The density ratio is defined as the difference between iron and silicate densities divided by the silicate density. In a real MO, this ratio is about $1.08$ (Table \ref{tab:table1}). In our models, we assign iron and silicate densities of $1.1$ and $1$, respectively, following the standard LBM convention of normalizing densities to unity. This results in an iron-silicate density ratio about one order of magnitude lower than that of a real MO. Such a reduction is necessary to maintain numerical stability and accuracy within the LBM framework, as large density differences can lead to spurious velocities and numerical instabilities in multiphase simulations. By restricting the density ratio to values near unity, we ensure stable and reliable convergence of our model.

To address this discrepancy, we adjust the model's gravity using the thermal parameters as described in Section \ref{sec:method}, by matching the ratio of compositional to thermal buoyancy forces in both model and MO. Specifically, the expression $(\frac{\Delta \rho g_{comp}}{\rho_{si}g_{conv}\beta\Delta T})_{\ell} = (\frac{\Delta \rho g}{\rho_{si}g\beta\Delta T})_{p}$ preserves the intended buoyancy ratios and flow behavior while remaining within the stability limits of our numerical scheme. In other words, we introduce a compositional gravity term ($g_{comp,\ell}$) to govern iron-silicate separation, using the model’s convective gravity ($g_{conv,\ell}$) and the values of the thermal expansion ($\beta$) and temperature difference in the model and real MO, as outlined below:

\noindent
\begin{equation}
    g_{comp,\ell} = (\frac{\rho_{si,\ell}\Delta\rho_{p}\beta_{\ell}\Delta T_{\ell}}{\rho_{si,p}\Delta\rho_{\ell}\beta_{p}\Delta T_{p}})g_{conv,\ell}
\end{equation}
\noindent
Here, $p$ refers to the MO parameters, and $\ell$ denotes the lattice units. Note that these thermal parameters are merely used to scale the compositional gravity properly, and we do not model thermal convection in this study.

The densimetric Froude number ($\text{Fr}$), which measures the ratio of inertial to gravitational buoyancy forces, is defined as  
$\text{Fr} = \frac{u}{\sqrt{g_{comp}\,(\Delta\rho / \rho_{\mathrm{si}})\,l}}$
where $\rho_{\mathrm{si}}$ is the silicate density. For large planetary impacts, $\text{Fr}$ values typically lie in the range $1\!-\!500$ \citep{allibert2023planetary}. Based on the model parameters listed in Table~\ref{tab:table1}, the Froude numbers in our simulations range from $0.08$ to $2.2$, reflecting conditions where inertial and buoyancy forces are of comparable magnitude.

\subsection{Timescale}
\label{sec:timescale}

Matching the $\text{Fr}$ formula (Section \ref{sec:model setup}) for the model and MO and using $u_p=u_{\ell}\frac{\Delta x_p}{\Delta t_p}$, where $\Delta x_p = L_p/L_{\ell}$, we define the physical time step as follows: 

\begin{equation}
     \Delta t_{p} = \sqrt{\frac{\Delta\rho_{\ell}\ g_{comp,\ell} \ \rho_{si,p}}{\Delta\rho_{p} \ g_{p} \ \rho_{si,\ell}} \Delta x_p}
     \label{eq:timescale}
\end{equation}
\noindent
$\Delta t_p$ is used to convert the model's dimensionless time to real time. Using Table \ref{tab:table1}, one modeling time step is equivalent to $0.0086\;\text{sec}$.

\section{Results}
\label{sec:result}

In this work, we investigate the effect of the impactor size and aspect ratio on the entrainment coefficient and iron-silicate mixing. In the following sections, we present the entrainment coefficients, the impactor's fragmentation within the MO, and explore the extent of iron-silicate mixing using the mixing entropy metric.


\begin{table}[htbp]
\centering
\begin{threeparttable}
\caption{MO and simulation parameters}
\label{tab:table1}
\small 
\begin{tabular}{|p{3.8cm}|c|c|p{4.5cm}|}
\hline
\textbf{Parameters} & \textbf{Magma ocean} & \textbf{Dimension} & \textbf{LBM} \\ \hline
Silicate density ($\rho_{Si}$) & $3750^{(1)}$ & $kg\;m^{-3}$ & $1$ \\ 
Iron density ($\rho_{Fe}$) & $7800^{(1)}$ & $kg\;m^{-3}$ & $1.1$ \\ 
Density ratio ($\Delta\rho/\rho_{Si}$) & $1.08$ & --- & $0.1$ \\
Gravity ($g$) & $9.8$ & m s$^{-2}$ & $10^{-3}$ (conv.), \newline $1.059 \times 10^{-5}$ (comp.) \\ 
Depth ($L$) & $3\times{10^6}^{(2)}$ & $m$ & $4095$ \\
Dynamic Viscosity ($\eta$) & $10^{-3}-1^{(3)}$ & $Pa.s$ & $2.67\times10^{-4}$ \\
Kinematic viscosity ($\nu$) & $10^{-7}-10^{-4}$ & $m^{2}s^{-1}$ & $8.28\times10^{-4}$ \\

Weber number ($\text{We}$) & $\geq{10^{14}}^{(4)}$ & --- & $567\text{--}894$ \\
Bond number ($\text{Bo}$) & $\geq{10^{14}}^{(4)}$ & --- & $99\text{--}1299$ \\\hline
\end{tabular}
\begin{tablenotes}
\item[1] Rubie et al. (2003); \item[2] Solomatov (2015); \item[3] Bajgain et al. (2022); \item[4] Deguen et al. (2014)
\end{tablenotes}
\end{threeparttable}
\end{table}


\subsection{Entrainment coefficient}

\begin{sidewaysfigure}
\centering
\includegraphics[
    trim=3 3 3 3,
    clip,
    width=1.03\textheight,
    height=0.8\textwidth,
    keepaspectratio
]{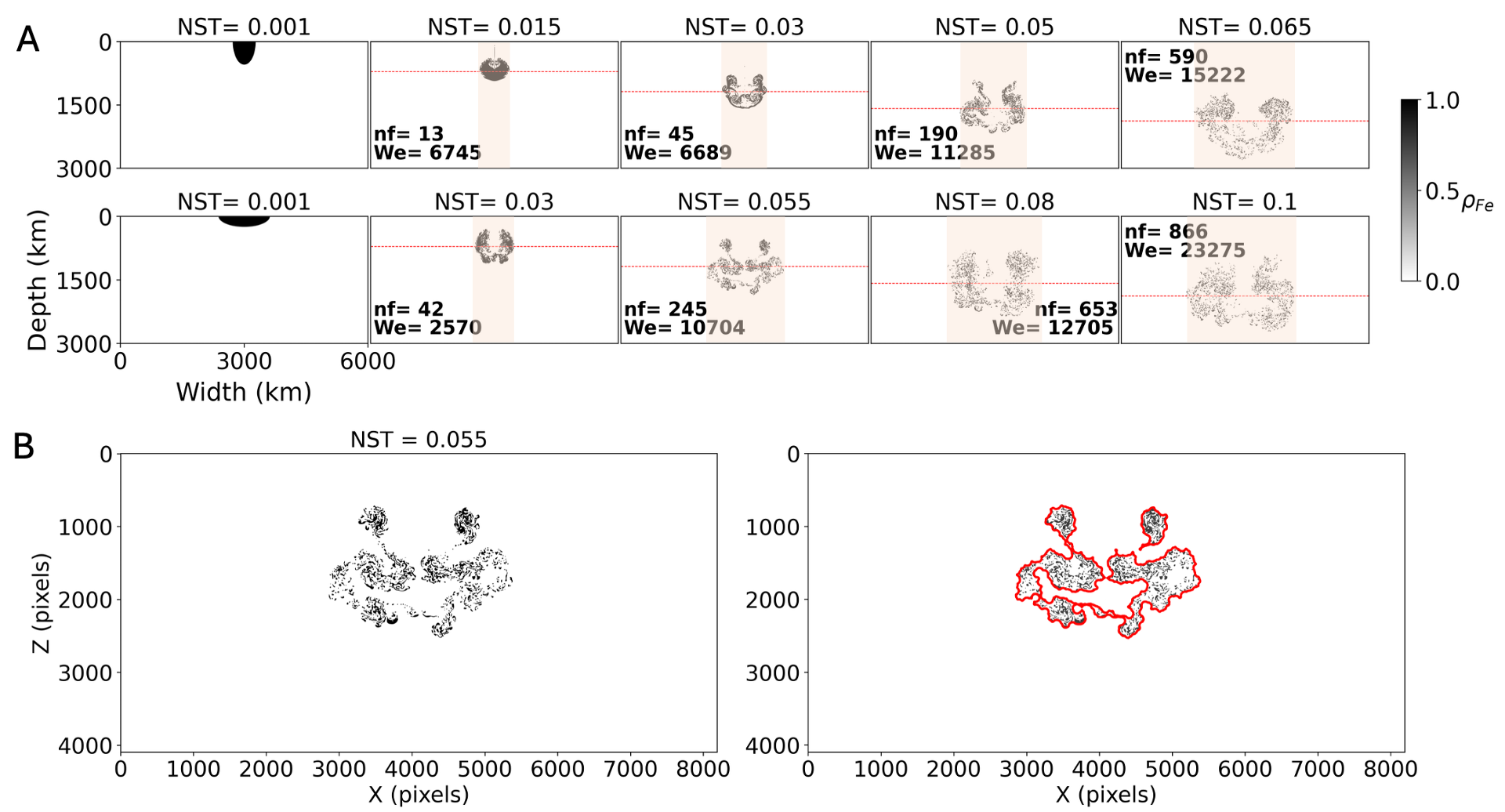}
\caption{Panel A shows the impactor evolution with a diameter of $552\;\text{km}$ and two different aspect ratios, $1$ (top) and $5$ (bottom), in the MO. The red dashed line is computed as the average vertical position of the iron cloud center of mass across snapshots for all aspect ratios in a given column. The shaded regions indicate the full extent of the iron cloud evolution in each panel. Panel B shows the plots of the iron cloud from an impactor with a diameter of $552\;\text{km}$ and an aspect ratio of $5$ (left), alongside the envelope encloses the iron cloud shown in red (right). Two red dots indicate the region within one standard deviation of the iron concentration distribution.}
\label{fig:fig2}
\end{sidewaysfigure}

Numerous studies have explored the deformation and fragmentation of an impactor throughout its interaction with the MO \citep{deguen2011experiments,deguen2014turbulent, landeau2014experiments,lherm2018small}. Likewise, our fluid dynamic numerical experiments exhibit fragmentation of the impactor and the development of an iron cloud controlled by variations in the impactor's morphology arising from velocity fluctuations within the turbulent MO \citep{villermaux2009single, tritton2012physical}. Figure \ref{fig:fig2}A illustrates the temporal evolution of an impactor with a diameter of $552\;\text{km}$ in the MO, shown for aspect ratios of $1$ (top panel) and $5$ (bottom panel). Simulation time is normalized by the total steps ($4\times10^{6}$; normalized simulation time, NST). The impactor is initialized at the top of the domain and sinks into a low-viscosity silicate phase. The sequence observed here for both aspect ratios mirrors the stirring regime reported in prior works \citep{ wacheul2014laboratory, lherm2018small, kriaa2024influence}, with the formation of sheets and ligaments that subsequently break into smaller fragments. This indicates a qualitative agreement between our results and previous works.

Visual inspection of the plots in Figure \ref{fig:fig2}A reveals a wider lateral expansion and a larger number of iron fragments for the impactor with a larger aspect ratio (second-to-last snapshots in the bottom panels). To obtain a quantitative insight, first, we defined a red dotted reference line in each panel to examine the impactor's evolution around it. This red line is determined by averaging the vertical coordinates of the iron cloud's center of mass in both snapshots for a given column with different aspect ratios. The orange shaded regions mark the range of the spreading fronts of the iron cloud in each panel. We used the scikit-image Python library \citep{scikit-image} to identify the iron fragments, the centroid and radius of each fragment, and the total number of fragments, $\text{nf}$, in each snapshot. These quantities are required to compute the local $\text{We}$, in every frame. For each snapshot, we estimated the surface tension in lattice units using the mode of the fragment radii (\textit{i.e.}, the most frequent radius in each frame) as described in Section \ref{sec:model setup}. Using the velocity field at the fragment centroid, together with the fragment radius, the inferred surface tension, and the model iron density, we calculated the corresponding $\text{We}$ value for each panel in Figure \ref{fig:fig2}A. As the impactor sinks deeper into the MO, the shaded regions in all these panels gradually expand, reflecting enhanced interaction between the impactor and the MO, an increase in the number of fragments, and larger local $\text{We}$ values. However, the impactor with the smaller aspect ratio undergoes less time-dependent evolution, as indicated by the smaller shaded area, and produces fewer iron fragments along the reference line than the impactor with the larger aspect ratio. Evidently, as expected, the temporal variations in $\text{We}$ control the deformation and breakup of the impactor \citep{pilch1987use, qaddah2019dynamics, qaddah2020thermal}. 

Using the entrainment hypothesis and following \citet{deguen2011experiments, deguen2014turbulent}, we have determined the entrainment coefficient, $\alpha$, for the impactor's evolution within the MO across all tested diameters and aspect ratios, as described below.

Figure \ref{fig:fig2}B, left panel shows a snapshot of the iron cloud of an impactor with a diameter of $552\;\text{km}$ and an aspect ratio of $5$ at NST = $0.055$. 

We employed the scikit-image and SciPy Python libraries to track the geometric evolution of the iron cloud. The right panel of Figure \ref{fig:fig2}B presents the envelope around the iron cloud, which is established by applying a morphological binary closing filter (\texttt{skimage.morphology}) with an optimized $50$-pixel circular structuring element, followed by a topological hole-filling operation (\texttt{scipy.ndimage}). The total cross-sectional envelope area, $\text{A}$, is calculated directly as the spatial sum of this solid mask, and its outer boundary perimeter, $\text{p}$, is extracted using the Marching Squares algorithm via \texttt{skimage.measure.find\_contours} at an isolevel of 0.5. To determine $\alpha$, we used the following relation \citep{morton1956turbulent}, across the simulation snapshots:
\noindent
\begin{equation}
    dA/dz = p\alpha
\end{equation}
and 
\begin{equation}
    \alpha = \frac{A-A_0}{\int{pdz}}
\end{equation}
\noindent
where $A_{0}$ is the initial cloud's envelope area. The vertical displacement step between successive snapshots, $\text{dz}$, is evaluated using \texttt{np.diff} on the vertical coordinate of the cloud’s center of mass. The cumulative spatial integral $\int{pdz}$ is then computed by accumulating the product of p and this step size via \texttt{np.cumsum}. After excluding the data points from the beginning and end of the trend that deviate from the linear growth of the iron cloud, which occurs primarily for smaller impactors with diameters of $152\;\text{km}$ and $252\;\text{km}$, a linear regression of the total envelope area, $\text{A}$, against $\int{pdz}$ is performed via \texttt{scipy.stats.linregress}. The slope of this linear fit yields $\alpha$. Figure \ref{fig:alphas},  panels A to E, show the corresponding plots for all diameters and aspect ratios.

To isolate the true entrainment behavior from edge artifacts, we initially placed the impactor at $500$ pixels ($366\;\text{km}$) below the upper boundary of the simulation domain and halted the cloud tracking at $500$ pixels above the lower boundary. However, mainly for the smaller impactor diameters ($152\;\text{km}$ and $252\;\text{km}$), we manually truncated the data points at the beginning and end of the time series to isolate the linear growth phase of the cloud. Despite this rigorous boundary curation, these smaller impactors are heavily influenced by the background MO flow, obscuring a completely clean linear growth pattern (Figure \ref{fig:alphas}, panels A and B) when compared to the larger impactor cases (Figure \ref{fig:alphas}, panels C to E). 

To evaluate the sensitivity of $\alpha$ to the initial conditions in our models, we conducted several simulations across all impactor diameters using a subset of aspect ratios, with the impactor placed at a depth of $500$ pixels below the surface and its horizontal position shifted by one grid cell. We calculated $\alpha$ for this set of simulations. Then we combined all alpha values from both models (with and without perturbations) into a single visualization for comprehensive statistical analysis.

\begin{sidewaysfigure}
\centering
\includegraphics[
    trim=3 3 3 3,
    clip,
    width=1.03\textheight,
    height=0.8\textwidth,
    keepaspectratio
]{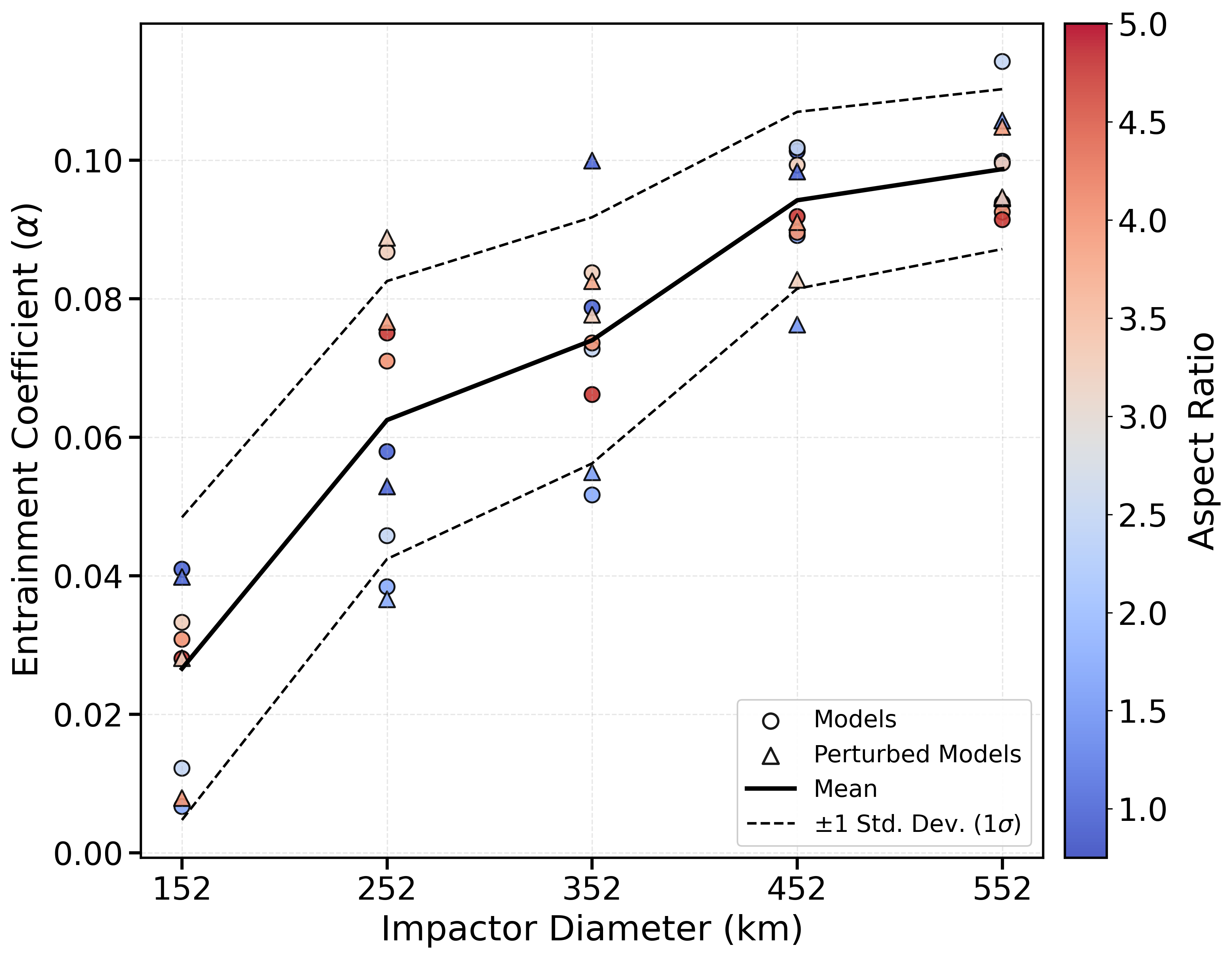}
\caption{Entrainment coefficient variation across different diameters and aspect ratios, with the mean (solid line) and standard deviation (dashed line).}
\label{fig:alpha_plot_final}
\end{sidewaysfigure}

Figure \ref{fig:alpha_plot_final} displays these combined datasets across all investigated impactor diameters and aspect ratios, utilizing circles for the models without perturbation and triangles for the perturbed models. To minimize visual crowding and clearly reveal any systematic trends in the color shading, the individual runs are visually simplified by averaging every three neighboring aspect ratios (\textit{e.g.}, binning aspect ratios $0.75$, $1$, and $1.25$ into an aspect ratio bin centered at $1$).The statistical trend lines including, the solid mean line, and the dashed standard deviation line are computed using an unbinned dataset.

The plot demonstrates that $\alpha$ increases systematically with increasing impactor size, falling within the range of $[0.00302, 0.12426]$ for models without perturbation and $[0.00723, 0.12240]$ for models with perturbations. Despite the slight variations in $\alpha$ for a given diameter across different aspect ratios, and the fact that these ranges are lower than the experimental values ($0.2-0.3$) determined by \citep{deguen2014turbulent}, likely due to dimensional constraints in our 2D simulations, the central finding remains robust, demonstrating a clear dependence of $\alpha$ on impactor diameter.

We also conducted a simulation for a large impactor with a diameter of $1100\;\text{km}$ and an aspect ratio of $5$, through which, we determined $\alpha= 0.07807$. When compared against our intermediate-sized cases, at the same aspect ratio, this value indicates that the entrainment coefficient may plateau or even begin to decrease at very large scales.


\begin{sidewaysfigure}[htbp]
    \centering
    \vspace{0.1cm} 
    
    \begin{minipage}[c]{0.53\textwidth}
        \centering
        \includegraphics[width=\textwidth]{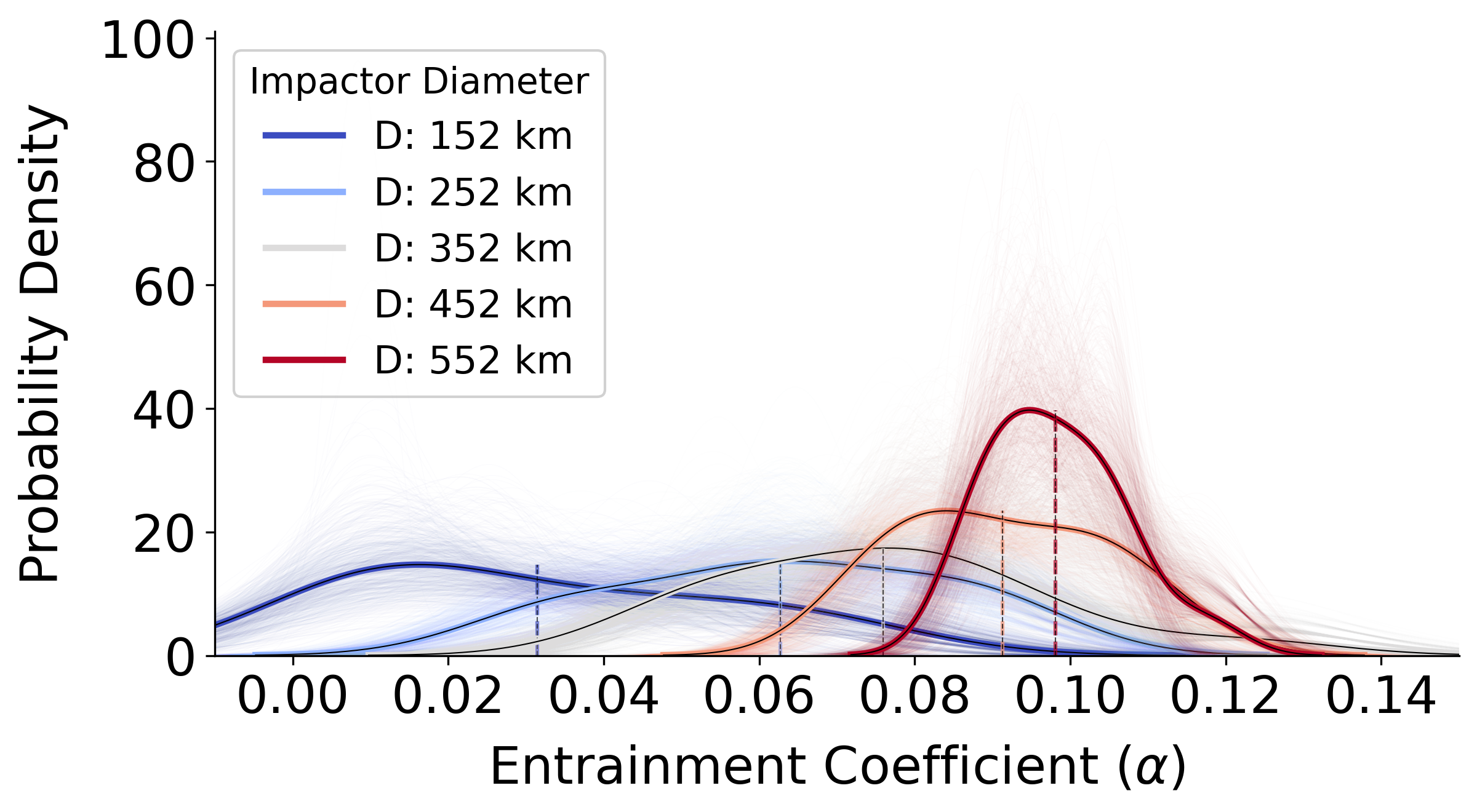}
    \end{minipage}
    \begin{minipage}[c]{0.43\textwidth}
        \centering
        \small 
        \begin{tabular}{cccc}
            \toprule
            \textbf{$D$ (\text{km})} & $\boldsymbol{\alpha_{\text{mean}}}$ & \textbf{95\% CI Lower} & \textbf{95\% CI Upper} \\ 
            \midrule
            152 & 0.031 & 0.018 & 0.045 \\
            252 & 0.063 & 0.052 & 0.073 \\
            352 & 0.076 & 0.066 & 0.087 \\
            452 & 0.091 & 0.085 & 0.098 \\
            552 & 0.098 & 0.094 & 0.103 \\
            \bottomrule
        \end{tabular}
    \end{minipage}
    
    \vspace{0.5cm} 
    \caption{The probability density functions for the bootstrapped entrainment coefficient. The solid colored curves represent the primary KDE, while the surrounding ensemble of $10000$ faint curves represents the bootstrapped distribution, highlighting the statistical variance. The colored dashed lines denote the expected mean $\alpha$ for each impactor diameter. The corresponding tables provide the mean $\alpha$ and the $95\%$ Confidence Intervals for each impactor diameter.}
    \label{fig:alpha_bootstrap_final}
\end{sidewaysfigure}

To further constrain the relationship between the impactor size and $\alpha$ despite finite sample sizes in both models with and without perturbation, we employed Kernel Density Estimation (KDE) combined with bootstrapping. To perform an unbiased bootstrapping, we combined the $\alpha$ values for a given impactor's diameter and identical aspect ratios from both models. By repeatedly resampling the data ($10000$ iterations), we generated probability density functions (PDFs) and calculated $95\%$ confidence intervals (CI) to define the likely range of the true physical average of $\alpha$. As illustrated by the probability density functions in Figure \ref{fig:alpha_bootstrap_final}, the ensemble mean of $\alpha$ increases systematically with impactor diameter, shifting from $\approx 0.031$ for the smallest body to $0.09$ for the largest, a trend that persists regardless of initial grid-scale perturbations. Smaller impactors ($D = 152\text{ km}$ and $252\text{ km}$) display broader, shallower curves, demonstrating that entrainment efficiency at these scales is highly dependent on the initial geometry of the impactor. Conversely, for larger impactors ($452\text{ km} \le D \le 552\text{ km}$), the curves become remarkably narrow and sharp, indicating that $\alpha$ becomes highly predictable at larger scales and rendering the dynamics completely insensitive to initial impactor's geometry.


\subsection{Fragmentation}
\label{sec:frag}
One key question in iron-silicate segregation dynamics concerns how efficiently the impactor’s iron core breaks up and whether it fragments into the smallest fragment size during its interaction with the MO before reaching the planetary core. This has important implications for iron-silicate equilibration efficiency and for the timescale of core formation \citep{nimmo2015early}. Here, we evaluate this scenario using our models, focusing on the fragmentation of the impactor within the MO.

\begin{sidewaysfigure}
    \centering
    \includegraphics[
        trim=5 5 5 5,
        clip,
        width=1.0\textheight,
        height=0.4\textwidth, 
        keepaspectratio
    ]{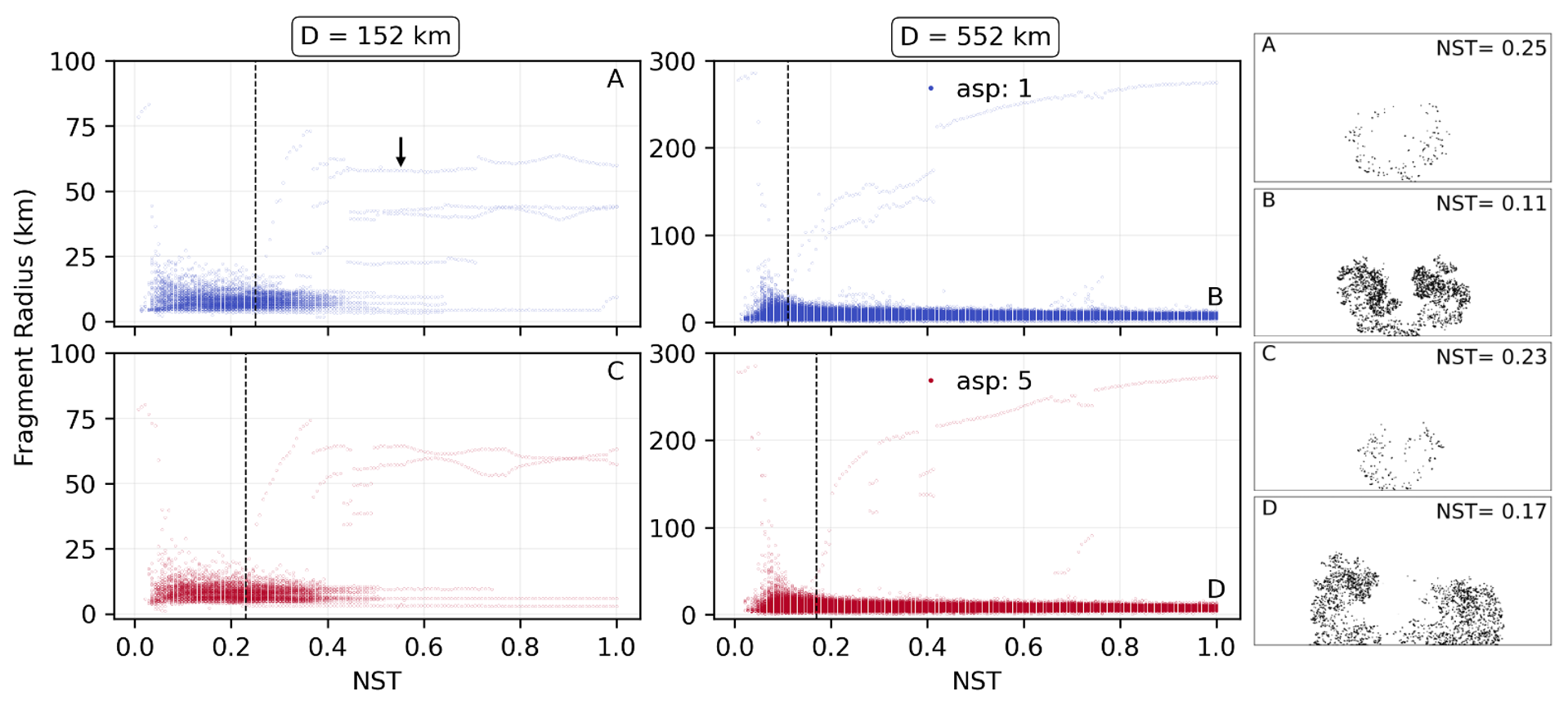}
    
    \vspace{0.5cm} 
    
    \includegraphics[
        trim=5 5 5 5,
        clip,
        width=1.0\textheight,
        height=0.4\textwidth,
        keepaspectratio
    ]{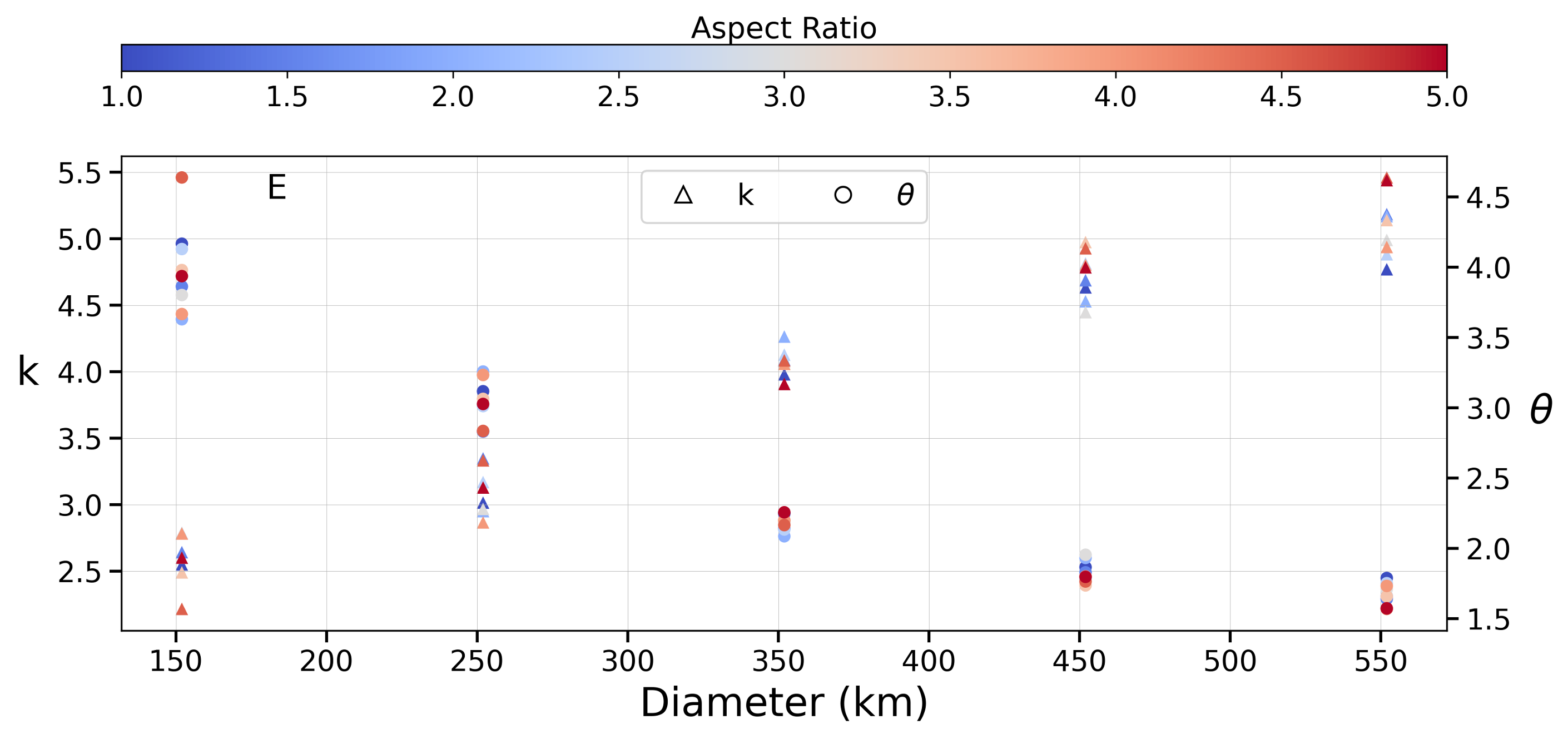} 
    
    \caption{Left panels (A–D) show the fragmentation process for impactor cores with diameters of $152\;\text{km}$ and $552\;\text{km};$, and aspect ratios of $1$ and $5$. The black arrow highlights an instance of either large sedimented layers or coalesced iron fragments. Black vertical dashed lines mark the times just before the impactor first arrives at the bottom of the MO. Right panels (A–D) depict the moments when the leading front of the iron cloud reaches the base of the MO, corresponding to the dashed lines on the left plots. The plot of $k$ and $\theta$ versus impactor diameters is shown in panel E, with different colors representing various aspect ratios.}
    \label{fig:fragment_onset_final}
\end{sidewaysfigure}

Our results show the impactor fragments into progressively smaller pieces through the interaction with the turbulent MO before reaching the MO base. Figure \ref{fig:fragment_onset_final}, Left panels (A-D) illustrate this process (derived from post-processing of simulation results) for two end-member cases of the impactor's diameter and aspect ratio: diameters of $152\;\text{km}$ and $552\;\text{km}$, each with aspect ratios of $1$ and $5$. These plots show the temporal evolution of fragment radii, with iron breaking into smaller fragments, reaching model-resolved sizes consistently less than $10\;\text{km}$. The trend highlighted by the black arrow in panel A (left panel) indicates either large iron layers settling at the MO base or coalesced larger fragments. The black dashed lines denote the time just before the iron cloud first arrives at the bottom of the MO. 

Right-side plots (panels A-D) show snapshots of the iron cloud at times marked by the black dashed lines on the left plots. Comparing left and right plots further confirms that fragments reach the smallest grid-resolvable size before the iron cloud arrives at the MO base. If domain resolution were increased to capture the turbulence of the real MO, we would expect fragmentation to continue to correspondingly the smallest scales before base arrival. Our results, as well as this expectation, suggest that MO dynamics provide sufficient time to fragment impactors extensively before core accumulation.

Our results also demonstrate that iron fragmentation yields a range of fragment sizes rather than a single characteristic radius, consistent with previous numerical and experimental studies \citep{villermaux2009single, ichikawa2010direct, wacheul2014laboratory}.  The evolution of the size distribution, where larger fragments break down into smaller ones through interactions with turbulent flow, is well described by the gamma function, $\Gamma(k)$, in the following equation:
\begin{equation}
    p(x) = \frac{x^{k-1}e^{\frac{-x}{\theta}}}{\theta^{k}\Gamma(k)}
\end{equation}
\noindent
where $x$ refers the any arbitrary radius on the distribution, $k$ denotes the shape of the gamma distribution, $\theta$ is the gamma function scale, and $\Gamma(k)$ presents the gamma function at $k$ value which is defined as $\Gamma(k)=(k-1)!$. Smaller $k$ and larger $\theta$ correspond to a broader, more variable size distribution, while larger $k$ and smaller $\theta$ yield a narrower, more uniform size distribution. By determining $k$ and $\theta$ across various impactor sizes, Figure \ref{fig:fragment_onset_final}E reveals a systematic dependence on diameter: $k$ increases while $\theta$ decreases as the impactor diameter grows, a trend that remains consistent across all investigated aspect ratios. These results indicate that smaller impactors (lower $k$) produce a greater diversity of fragment sizes, whereas larger impactors (higher $k$) yield more similar-sized fragments. This diversity may enhance mixing efficiency, since smaller fragments are surrounded by more silicate in a given volume of the MO \citep{deguen2014turbulent, landeau2021metal}, while the more uniform fragment sizes associated with larger impactors may reduce the overall mixing efficiency.


\begin{sidewaysfigure}
\centering
\includegraphics[
    trim=5 5 5 5,
    clip,
    width=1.03\textheight,
    height=0.8\textwidth,
    keepaspectratio
]{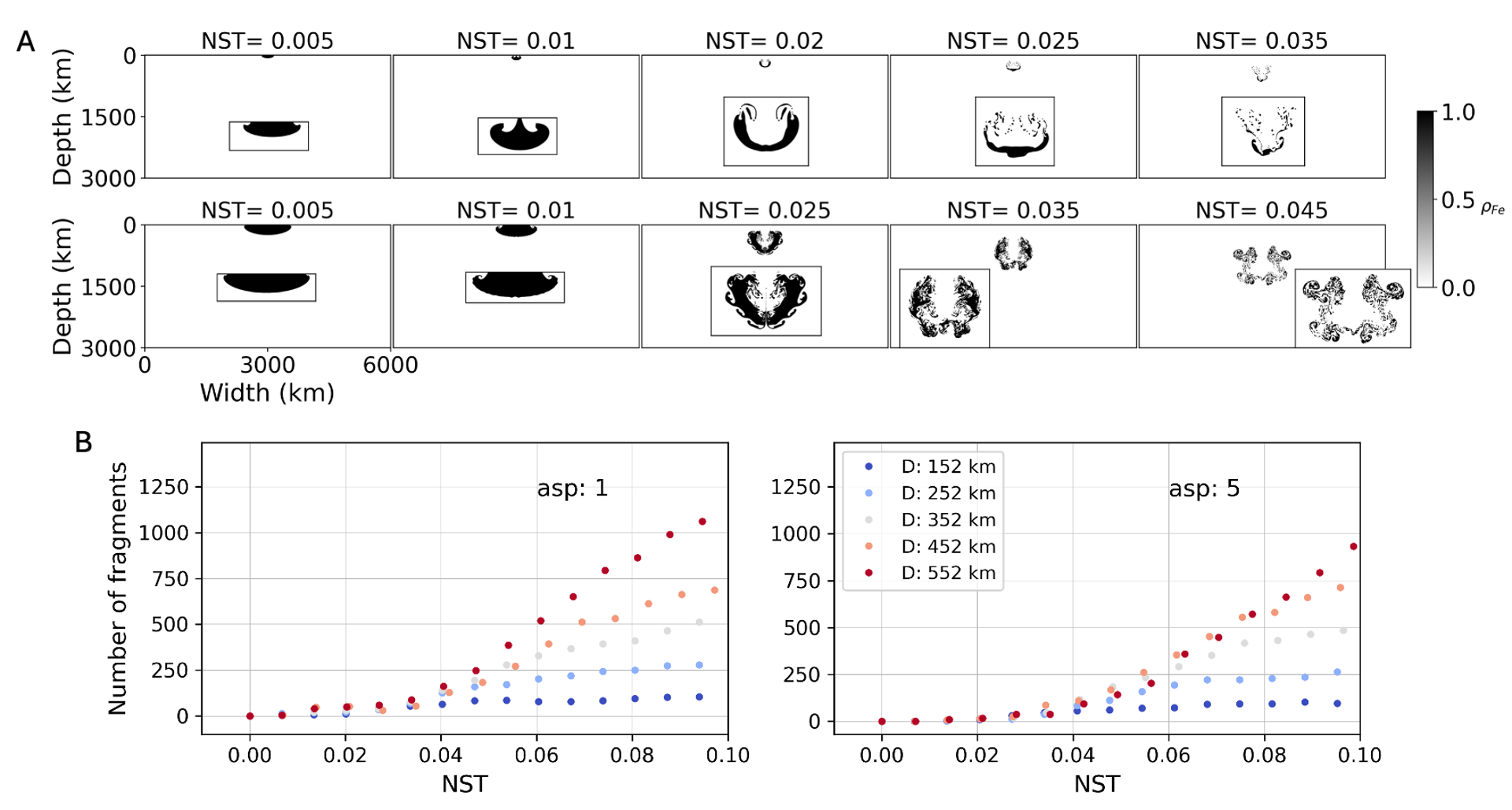}
\caption{Panel A shows the successive fragmentation of impactors with diameters of $152\;\text{km}$ (top) and $552\;\text{km}$ (bottom), both with an aspect ratio of $5$. Insets depict the magnified impactor evolution in each frame, for better visualization. Panel B shows the number of fragments versus NST for all impactor diameters, at aspect ratios of $1$ (left) and $5$ (right).}
\label{fig:fragmentation_onset_start_final}
\end{sidewaysfigure}

Whether core fragmentation occurs as a single, global event \citep{deguen2014turbulent} or through a cascading process \citep{rubie2003mechanisms, dahl2010turbulent, samuel2012re} remains a pivotal question, as the specific mechanism dictates the available time and surface area for chemical exchange. A single-event model implies that equilibration is governed by the initial breakup geometry, whereas a cascading process suggests a continuous creation of new interfaces, allowing for a much higher degree of metal-silicate interaction as the iron descends through the MO. Consistent with a previous study by \citet{deguen2014turbulent}, our results show that the impactor first entrains and incorporates ambient silicate before fragmentation starts. Figure \ref{fig:fragmentation_onset_start_final}A demonstrates the successive fragmentation of the impactor with diameters of $152\;\text{km}$ (top panels) and $552\;\text{km}$ (bottom panels), both with an aspect ratio of 5. Insets in each panel enlarge the impactor’s evolution at each stage to better visualize the fragmentation process. As evident for the smaller impactor, fragmentation initiates at the upper parts of the elongated core at NST= 0.02 and progresses over time, whereas for the larger impactor it starts at the bottom edge (Rayleigh-Taylor-type erosion at the iron-silicate interface, as proposed by \citet{dahl2010turbulent}) at NST= 0.01 and proceeds with well-defined fragments over time. Our results also confirm that this progressive breakup is consistent with the fragmentation cascades described by \citet{rubie2003mechanisms} and \citet{samuel2012re}. In other words, in contrast to the predominantly single fragmentation episode inferred by \citet{deguen2014turbulent}, our simulations show a cascading fragmentation sequence over NST= $0.02-0.035$ for the smaller impactor and NST= $0.01-0.045$ for the larger impactor, continuing until the smallest resolved fragment sizes are reached (not shown here).

Our models reveal that impactors of different diameters begin fragmenting at comparable times. Figure \ref{fig:fragmentation_onset_start_final}B demonstrates the onset timing of fragmentation by showing the number of fragments as a function of NST for all impactor diameters, for aspect ratios of $1$ (left panel) and $5$ (right panel). Early portions of these curves highlight size- and aspect ratio-independent fragmentation onset. These results, which are driven from pure-turbulence models, match \citet{rohlen2025impactor} (their Fig. 5), who found similar fragmentation onset timing for different impactor sizes at fixed impact velocity and MO thickness. Their analysis (including a combination of impacts and turbulence dynamics) shows that smaller impactors form shallower craters while larger ones excavate deeper craters. This, along with the comparable fragmentation onset time in their results, implies that crater formation dissipates initial momentum such that MO turbulence governs the fragmentation across impactor sizes. This consistency between our results and theirs suggests that even giant impactors fragment to some extent during descent through sufficiently deep and turbulent MOs.

\subsection{Analysis of Iron-Silicate Interfacial Distribution}
\label{sec:iron-silicate mixing}
To constrain the iron-silicate interaction and mixing extent in our models, we analyzed the raw simulation density fields (stored as \texttt{.npy} arrays), in which each grid cell has one of these values: $0$ for pure silicate, $1$ for pure iron, and intermediate values between $0$ and $1$ representing the iron-silicate interface. We note that while a resolution gap exists between our planetary-scale models and a real MO\textemdash where a continuous hydrodynamic cascade eventually reduces the iron core down to centimeter-scale droplets\textemdash our models explicitly capture the large-scale hydrodynamic fragmentation regime. We propose that these resolved large fragments and the overall spatial distribution of the iron cloud define the interaction envelope in which the intermediate interface grid cells represent the specific regions of the MO where further sub-grid droplet breakdown would be concentrated. To quantify the extent of the iron cloud's spatial expansion and iron-silicate interface characteristics, we adopted the mixing entropy metric \citep{zhu2024study}, which is derived from the information entropy of a binary system (iron and silicate in this study). This metric is defined as the weighted average information content of the simulation grid, computed from the phase values assigned to each grid cell:

\begin{equation}
    ME = \frac{IE}{IE_{max}} = \frac{-\Sigma_{i=0}^{N}[p_{i}ln(p_{i})+(1-p_{i})ln(1-p_{i})]}{-N[ClnC+(1-C)ln(1-C)]}
\end{equation}

Here, $N$ is the total number of grid cells in the field, while $p_{i}$ and $1-p_{i}$ denote the probabilities of iron and silicate in the 
$i$th cell, respectively. The term $IE_{max}$ is a normalization factor, calculated based on the global iron concentration, $C$, in the MO, where $C$ is the ratio of the initial impactor’s area to the total MO area. This factor ensures that maximum mixing $ME=1$ is achieved when the iron cloud is uniformly distributed throughout the MO. The mixing entropy metric, $ME$, ranges from $0$ to $1$, representing insignificant cloud distribution to through cloud's distribution inside the MO. It is important to note that there is a significant scale separation between the meter-to-kilometer-scale features resolved in our simulations and true millimeter-to-sub-meter emulsification processes in the MO. In other words, this mixing entropy metric differs fundamentally from thermodynamic mixing at the atomic scale. $\text{ME}$ in this study serves as a macroscopic geometric proxy that quantifies the degree of spatial intermingling and fluid interface stretching at the grid scale. In a real MO, chemical equilibration is governed by the total available contact surface area between the immiscible iron and silicate phases. Under high shear stress, the descending iron mass stretches into thin filaments and sheets; when these structures become narrower than the local grid resolution in our simulations, the corresponding cells register intermediate phase ($0 < p_i < 1$), driving an increase in $\text{ME}$. Consequently, while an elevated $\text{ME}$ does not imply atomic-level dissolution, it quantitatively tracks the expansion of the fluid interface envelope and the spatial dispersion of the iron cloud. This effectively identifies the volume fraction of the domain where large-scale dynamics have optimized the conditions required for sub-grid droplet breakdown.
 
To analyze the mixing dynamics, we distinguish between instant mixing entropy, which captures the spatial dispersion state of the iron cloud at any specific point in time, and the cumulative time fraction of mixing entropy, which aggregates these instantaneous states across the entire simulation to evaluate the percentage of total run time the system spent below a given mixing threshold.

Our results reveal that the overall extent of mixing and mixing rate are relatively insignificant across all impactor sizes and aspect ratios. This behavior is qualitatively consistent with previous theoretical and experimental studies \citep{dahl2010turbulent, landeau2014experiments, deguen2014turbulent}, which predict incomplete emulsification for large impactors ($D \geq 10\text{ km}$) inside the MO. We note that while those studies evaluate incomplete mixing via droplet-scale fluid emulsification length scales, the low $\text{ME}$ values in this study reflect a limited large-scale (kilometer-sized) geometric dispersion of the iron cloud governed by the competition between turbulent entrainment and gravity settling.

\begin{sidewaysfigure}
\centering
\includegraphics[trim=5 5 5 5,clip,width=1.03\textheight, height=0.8\textwidth, keepaspectratio]{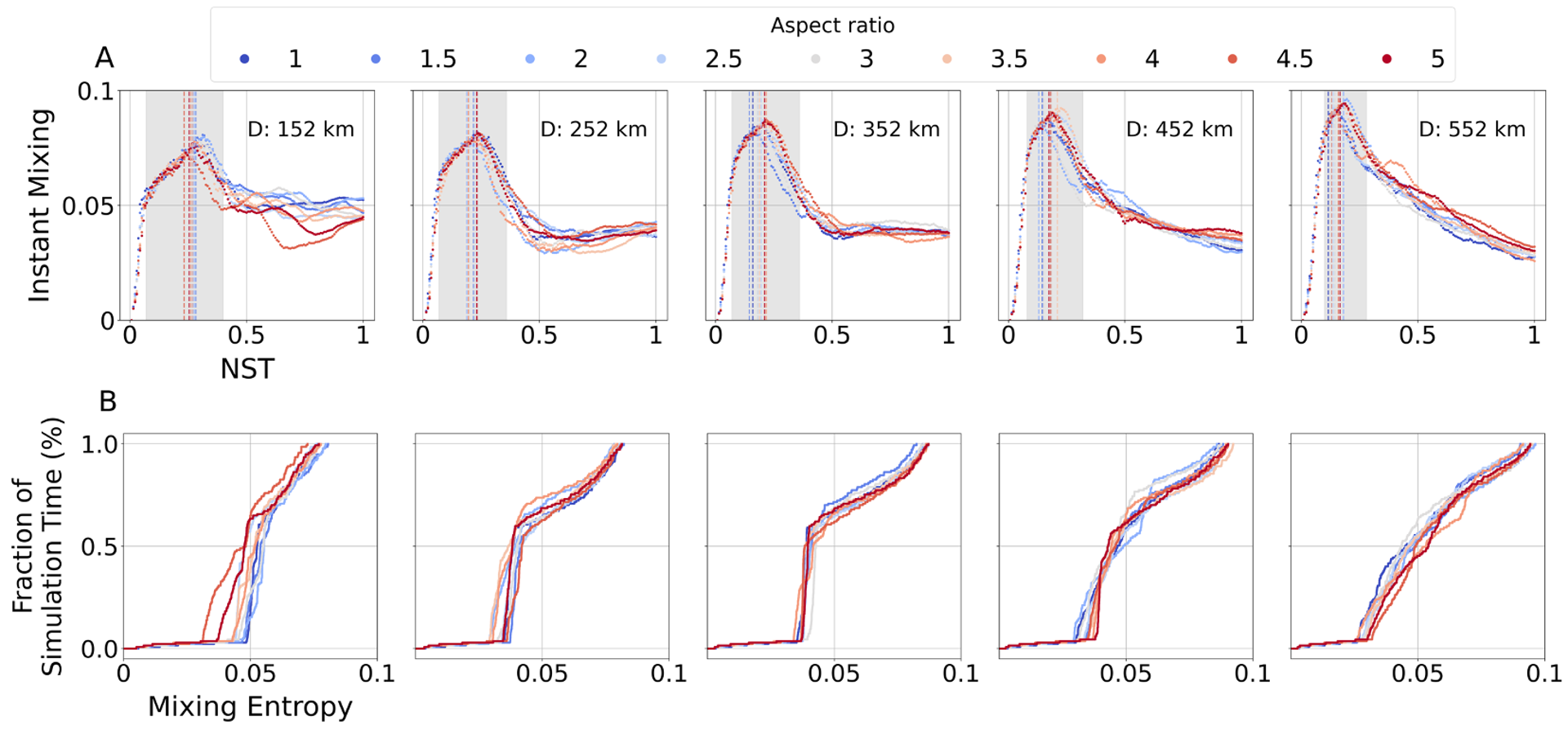}
\caption{Panels A and B show the plots of the instantaneous mixing and cumulative time fraction of mixing entropy, respectively, for impactors with increasing diameters from left to right. Colors indicate different aspect ratios, and shaded regions in panel B mark areas of maximum mixing in each case. The vertical dashed lines show the times at which the impactor initially has reached the bottom of the MO.}
\label{fig:mixing_4096_8192}
\end{sidewaysfigure}

Figure \ref{fig:mixing_4096_8192}A shows the instantaneous mixing entropy, tracking the dynamic evolution of the iron cloud's spatial dispersion throughout the simulation time. These plots demonstrate that the peak mixing interval (represented by gray shaded regions) increases with larger impactor sizes. In other words, as the impactor size increases, a larger portion of the iron phase penetrates regions of pure silicate within the MO. The vertical dashed lines denote the times at which the leading front of the iron cloud first reaches the base of the MO. Following these intervals, the decreasing trend of the curves marks the subsequent sedimentation of the bulk of the iron at the bottom of the MO, while the plateaus at higher NSTs are associated with the smaller iron fragments that remain dynamically suspended in the MO.

To explore the extent of cloud dispersion outside the primary size range investigated in this study, we evaluated the instantaneous mixing dynamics for extreme impactor diameters of $50\text{ km}$ and $1100\text{ km}$, both modeled with aspect ratios of $1$ and $5$. The $1100\text{ km}$ case represents a massive, protoplanetary-scale impactor. Because the RKLBM numerical algorithm destabilizes under severe local velocity gradients that exceed the low-Mach-number limit ($\text{Ma} > 1$), our simulation for the $1100\text{ km}$ impactor with an aspect ratio of $1$ suffered from numerical divergence and could not be completed. The temporal evolution of the instantaneous mixing for the successful simulations is shown in Figure \ref{fig:shannon_larger}. The iron cloud of these impactors exhibits limited geometric dispersion across the MO, which closely aligns with the mixing trends established in our intermediate-size impactor models.

To aggregate the temporal pathways shown in panel A, Figure \ref{fig:mixing_4096_8192}B illustrates the cumulative time fraction of mixing entropy, which explicitly quantifies the percentage of total simulation runtime the system spends at or below a given mixing threshold. For the smallest impactor ($D=152\;\text{km}$), the mixing entropy reaches a maximum of approximately $0.072$ at a cumulative time fraction of $100\%$. This maximum shifts to $0.094$ for the largest impactor ($D=552\;\text{km}$). The profiles in Figure \ref{fig:mixing_4096_8192}B reveal a steep vertical increase between $0.04$ and $0.05$ for diameters up to $352\;\text{km}$. This steepness indicates that the iron fragments reside within a narrow, nearly identical range of spatial dispersion states for the vast majority of the simulation history. In contrast, for larger impactors ($D \ge 452\;\text{km}$), the distribution shifts to the right and exhibits a lower and more gradual slope, signifying a wider variety of mixing states over the simulation time. While the maximum mixing entropy for the largest impactor is higher, this broader distribution indicates that the iron cloud exhibits a more diverse dispersion. Furthermore, the influence of the initial aspect ratio on the mixing entropy profiles becomes increasingly negligible as the impactor size increases.

Determining the spatial location of the iron cloud during the peak mixing intervals (shaded regions in Figure \ref{fig:mixing_4096_8192}A) provides critical insight into the depth of primary interaction of the iron cloud with the MO. Our results indicate that maximum primary mixing of the cloud with km-sized fragments is concentrated in the middle to deeper layers of the MO.

\begin{sidewaysfigure}
\centering
\includegraphics[trim=5 5 5 5, clip, width=0.9\textheight, height=0.9\textwidth, keepaspectratio]{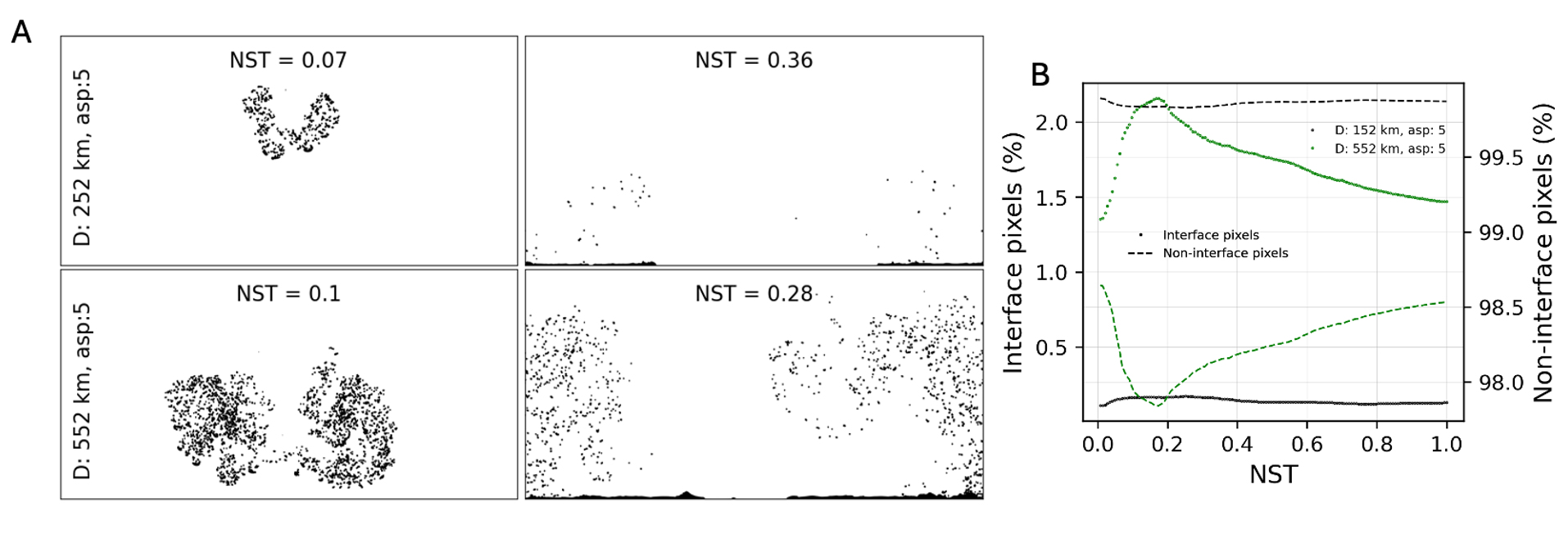}
\caption{Panel A shows the simulation snapshots of the impactor fragmentation with diameters $252\;\text{km}$ (top rows) and $552\;\text{km}$ (bottom rows), both with an aspect ratio of $5$. In each row, the snapshots are taken at the times corresponding to the lower and upper bounds of the shaded areas in Figure \ref{fig:mixing_4096_8192}B. Panel B presents the plot of the temporal evolution of the interface pixels (marked by a dotted line) and non-interface pixels (marked by a dashed line), for the impactor diameters and aspect ratio shown in panel A.}
\label{fig:mixing_image_4096_8192}
\end{sidewaysfigure}

Figure \ref{fig:mixing_image_4096_8192}A displays snapshots for $252\;\text{km}$ and $552\;\text{km}$ impactors at times corresponding to the lower and upper bounds of these peak mixing intervals, illustrating the transit of the iron cloud into the deeper mantle. These regions serve as proxies for where further breakups (not captured in this study) are potentially most intense. In a real MO, the smallest turbulent eddies scale as $d \propto Re^{-3/4}$ \citep{succi2001lattice}, implying that the extreme turbulence of a planetary-scale mantle would further break these resolved fragments into far smaller droplets \citep{rubie2003mechanisms, deguen2014turbulent} compared to what our numerical model can achieve, although the ultimate stable droplet size will be governed by the balancing effect of surface tension. Consequently, the mid- to lower MO layers identified in our models remain the probable zones for enhanced interface complexity and chemical exchange, provided that small-scale droplets (following further breakup) reach terminal Stokes velocity and their diffusion equilibration time is comparable to their residence time.

Figure \ref{fig:mixing_image_4096_8192}B tracks the temporal evolution of pixel categories for $152\;\text{km}$ and $552\;\text{km}$ impactors, at a fixed aspect ratio of $5$. The interface pixels (dotted lines) represent grid cells containing a mixture of phases, while non-interface pixels (dashed lines) represent pure iron or silicate. The number of interface and non-interface pixels in both cases is normalized by the total number of pixels in the MO. By increasing the impactor diameter, the proportion of the interface pixels increases, indicating a more extensive spatial scattering of the iron phase in the MO; however, the absolute mixing entropy remains limited (<$0.1$, Figure \ref{fig:mixing_4096_8192}A) across all sizes and aspect ratios.

\section{Discussion}
\label{sec:discussion}

We investigated the post-collisional evolution of the intermediate-sized differentiated impactor’s iron core within the MO, varying its size and aspect ratio.

At sufficiently high Reynolds numbers, our simulations capture the breakup of the impactor core into an iron cloud, followed by its fragmentation into kilometer-scale bodies. While our models do not resolve the small-scale sub-range where efficient metal-silicate equilibration occurs, these results characterize the large-scale dynamics of the fragmentation cascade. These early-stage behaviors are expected to be universal in high-Reynolds turbulent flows, as confirmed by laboratory experiments and hydrocode simulations operating at lower, but comparably inertial, Reynolds numbers than those of real MO \citep{deguen2011experiments, samuel2012re, deguen2014turbulent, kendall2016differentiated, landeau2014experiments, landeau2021metal, rohlen2025impactor}.

Our results for a fixed impactor diameter indicate that impactors with different shapes (aspect ratios) follow distinct evolutionary paths at a given MO depth (Figure \ref{fig:fig2}A), including systematic variations in the number of fragments produced during their interaction with the surrounding flow.
According to our findings, more elongated impactors with slightly different cloud dispersion geometry may contribute differently to mantle mixing and chemical modification. Specifically, more compact bodies produce fewer fragments and a less extended iron cloud, implying a reduced influence on material transport and mantle chemistry compared to their more elongated counterparts.

According to laboratory experiments by \citet{maller2024condition}, the Froude number plays a key role in the impactor fragmentation, with a transition from largely intact cores to complete fragmentation as $text{Fe}$ increases from below $10$ to around $40$ for single-phase impactors, depending on the impactor size and impact speed. In contrast, as in iSALE hydrocode simulations \citep{rohlen2025impactor}, which exhibit substantial fragmentation over a broad range of Froude numbers (from $9.17$ to $736$), our models also show extensive fragmentation but at much lower Froude numbers ($0.08$--$2.2$) than those explored in the laboratory experiments of \citet{maller2024condition}. While these previous studies define the Froude number exclusively at the moment of impact, they do not constrain the subsequent post-impact fragmentation dynamics. Our model fills this critical gap by tracking the long-term, post-impact fate of the impactor. Our results demonstrate that fragmentation continues for a significantly longer duration after the initial impact stage. This highlights that at the planetary scales of an MO, initial impact momentum is not the sole driver of core fragmentation; instead, buoyancy-driven instabilities can efficiently trigger and sustain fragmentation even in low-velocity, low-energy regimes.

Using the entrainment hypothesis, we find that the entrainment coefficient in our models ranges from $0.00302$ to $12426$, which is roughly one order of magnitude smaller than the experimental values ($0.2$--$0.3$) \citep{deguen2014turbulent}. This discrepancy can be attributed to several factors:
(i) The experimental setup employs a metal tube -- analogous to an impactor core -- with an aspect ratio of $1$, while our simulations explore a range of aspect ratios. (ii) The experimental MO has dimensions of $25.5 \times 25.5 \times 47\;\text{cm}^3$, corresponding to an aspect ratio of $0.54$, compared to the MO aspect ratio of $2$ used in this study.
(iii) Our simulations neglect compressibility effects associated with supersonic flows post-impact due to limitations of our methodology, which is not well-suited for modeling high-speed compressible dynamics.
(iv) The 2D geometry of our numerical models suppress 3D vortex stretching and prevent the turbulent energy cascade. Statistical analysis of our data indicates that $\alpha$ depends on the impactor's diameter (Figure \ref{fig:alpha_plot_final}). This size dependency is further validated by a bootstrapping analysis, which confirms a clear positive correlation where larger impactors yield higher entrainment coefficients (Figure \ref{fig:alpha_bootstrap_final}).

Our models do provide new insight into how much silicate is entrained in the downward flow associated with the descent of impactor-derived iron. Previous calculations of MO and core chemistry have assumed that entrainment coefficients are constant for a wide variety of impactors (e.g., \cite{rubie2015accretion}). The net impact of this assumption is that larger impactors are predicted to have little silicate entrainment, and therefore, the chemical effect of larger impactors on MO is muted. Within the range of impactor sizes examined in this study, we show that larger impactors have higher entrainment coefficients than smaller impactors, countering the geometric limitations that larger impactors have in entraining silicate and reacting with the MO. This finding carries profound geochemical implications, indicating that the chemical effects of larger impactors may be progressively enhanced rather than muted, as was previously thought. However, an exploratory simulation of a significantly larger planetary-scale impactor yielded an $\alpha$ value slightly lower than that of the largest body in our intermediate size range, suggesting that the entrainment coefficient may eventually plateau or decrease at extreme scales. Because this insight is drawn from a single exploratory run, these results highlight the need for a comprehensive, high-resolution study dedicated specifically to the large-scale impact scenarios to establish a firmer baseline.

Our results demonstrate a sequential fragmentation of the impactor into its smallest fragment size (permitted by the model resolution) before its first arrival at the bottom of the MO (Figure \ref{fig:fragmentation_onset_start_final}). This implies that in extremely turbulent MOs, where fragments reach scales $5$ to $6$ orders of magnitude smaller than those in our models, such fine droplets would settle much more slowly than our $\text{km}$-sized fragments. Longer residence times would enhance iron-silicate equilibration opportunities, provided thermodynamic and kinetic conditions are met. Therefore, relying on our models, one anticipates that turbulence of the real MO provides sufficient time for extensive fragmentation before core accumulation, creating favorable conditions for metal-silicate equilibration during planetary differentiation.

In this investigation, we focused on the post-impact physical processes. and did not model the impact dynamics themselves, including crater formation and jet release as reported in previous numerical studies using iSALE \citep{tonks1993magma, kendall2016differentiated, rohlen2025impactor} nor the chemical interactions and diffusion between iron and silicate. We also did not examine the MO immediately after impact, when shock heating is especially intense near the impact point \citep{nakajima2021scaling}. These aspects, which would strengthen and extend our findings and provide a more complete picture of iron-silicate segregation dynamics, will be addressed in future complementary research. Nevertheless, as an initial step in evaluating our models, it is important to perform simulations without thermal effects, so that they can be directly compared with laboratory experiments and with iSALE calculations, which likewise neglect thermal evolution and viscous diffusion at the scales considered here.


\section{Conclusion}
\label{sec:conclusion}

In this study, we investigated the post-collisional temporal evolution of a differentiated impactor's iron core as it interacts with the dynamics of a fully molten $3000\;\text{km}$-thick MO. We examined core diameters from $152$ to $552\;\text{km}$ (increments of $100\;\text{km}$) and aspect ratios from $0.75$ to $5$ (increments of $0.25$), assuming no thermal effects.

Our models demonstrate that MO turbulence strongly modifies the shape of the impactor core, fragmenting it into an iron cloud and kilometer-sized iron fragments dispersed throughout the MO. Fragmentation proceeds progressively through fragmentation cascade, generating additional fragments that continue to interact with the flow until they reach the grid resolution limit of our numerical models.

We determined the entrainment coefficients for various impactor diameters and aspect ratios, ranging from $0.00302$ to $0.12426$. Our analysis reveals a size-dependent relationship, where the entrainment coefficient increases with impactor diameter. By employing a bootstrapping method with $10000$ resamples on the entrainment data, we also found that larger impactors exhibit higher entrainment coefficients during their interaction with the MO. Our findings indicate that elongated impactors significantly enhance the iron-silicate interfacial area, thereby paving the way for further fragment breakups. While tracking these subsequent, sub-scale breakups explicitly is beyond the current resolution limits of our models, our results establish a critical baseline for understanding the initial stages of core emulsification. Using the mixing entropy metric, we found that iron cloud dispersion remains limited for all impactor sizes considered in this study, including the intermediate range, as well as the $50\;\text{km}$ and $1100\;\text{km}$ cases outside it. This restricted dispersion is in qualitative agreement with previous theoretical, numerical, and experimental work. Our results also indicate that the middle to lower layers of the MO exhibit a wider dispersion of the iron cloud. This creates favorable conditions for both metal-silicate equilibration and further fragment breakup, though capturing these subsequent sub-grid scale breakups remains beyond our current resolution limits.

\printcredits

\section*{Declaration of competing interest}
The authors declare that they have no known competing financial interests or personal relationships that could have appeared to influence the work reported in this paper.

\section*{Acknowledgements}
We would like to thank NASA for supporting this research through the NASA Emerging Worlds Grant 80NSSC21K0377. We thank the anonymous reviewers for their constructive comments, which greatly improved the quality of the manuscript.

\section*{Data Availability}
The dataset presented in this paper is generated from a combination of numerical models \citep{mora2021optimal, mora2018simulation, mora2023hpc} and is available in Mendeley Data (link: \url{https://data.mendeley.com/datasets/g4c6fsk79t/2}) and \citet{Honarbakhsh2026}.











\begin{appendices}

\section*{Appendix A} 
\counterwithin{figure}{section} 
\setcounter{figure}{0} 
\renewcommand{\thefigure}{A\arabic{figure}} 
\section*{Entrainment Coefficient Plots}
Here, we present the variations of the iron cloud's area plotted against the product of its perimeter and the change in its vertical center of mass. Figure \ref{fig:alphas}, panels A through E display these results across all aspect ratios for impactor diameters ranging from $152\;\text{km}$ to $552\;\text{km}$.

\begin{sidewaysfigure}
\centering
  \begin{subfigure}[b]{1.03\textheight}
    \centering
    \includegraphics[trim=5 5 5 5,clip,width=\textwidth,height=0.8\textwidth,keepaspectratio]{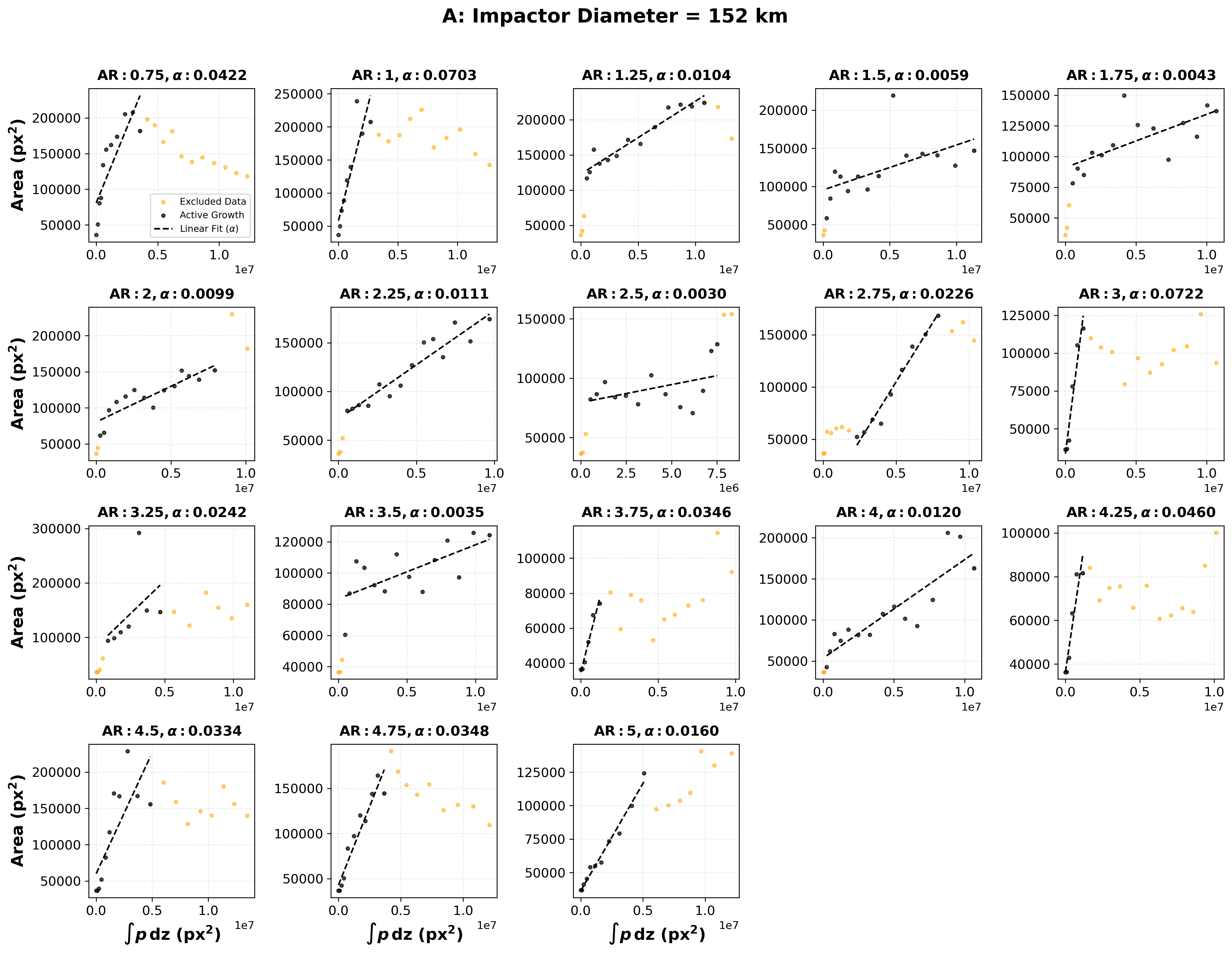}
    \label{subfig:alpha_152}
  \end{subfigure}
  \caption{Linear regression of the total envelope area, $A$, against $\int \rho dz$ across all aspect ratios (Panel A).}
  \label{fig:alphas} 
\end{sidewaysfigure}
\clearpage
\begin{sidewaysfigure}
\centering
\ContinuedFloat 
  \begin{subfigure}[b]{1.03\textheight}
    \centering
    \includegraphics[trim=5 5 5 5,clip,width=\textwidth,height=0.8\textwidth,keepaspectratio]{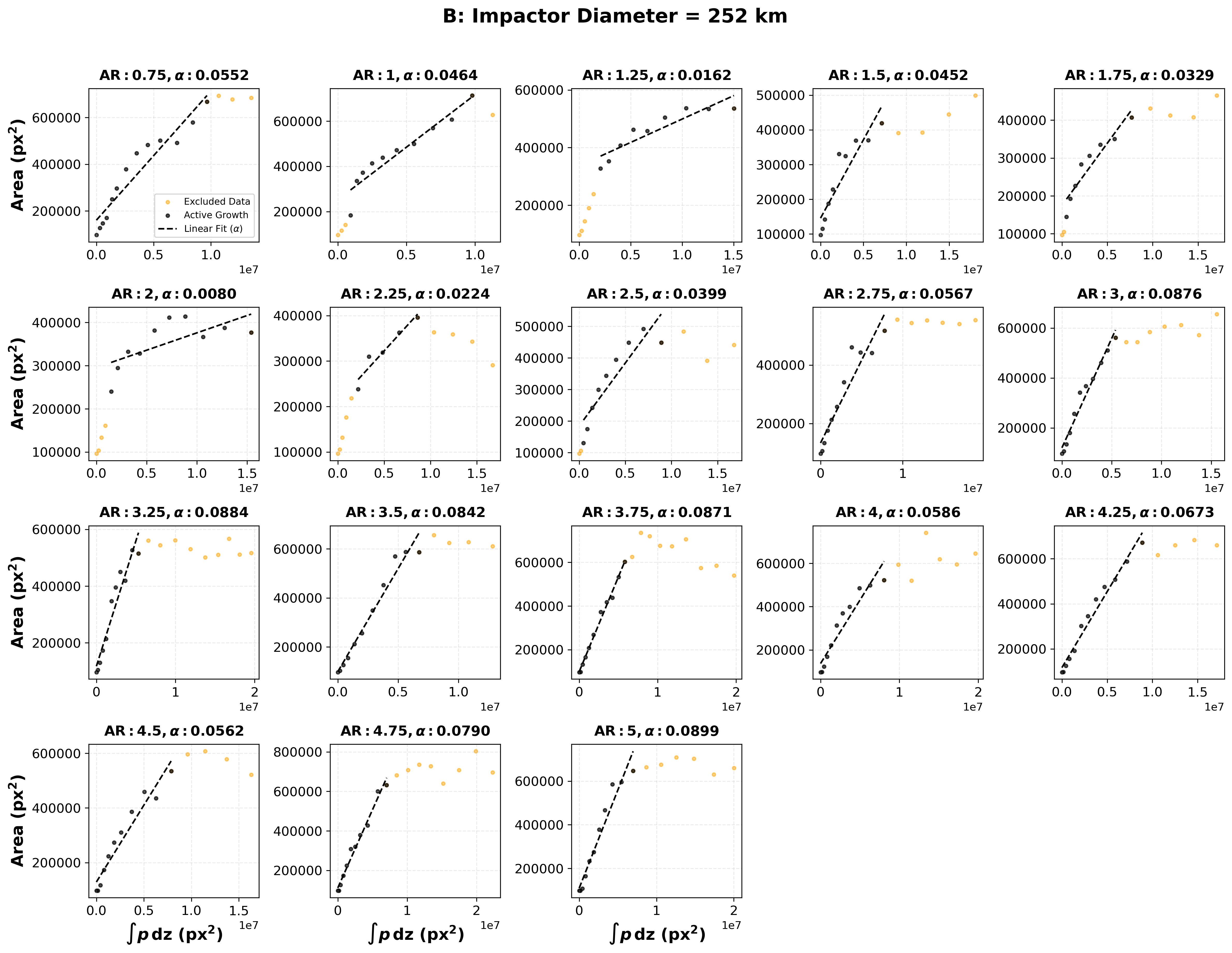}
    \label{subfig:alpha_252}
  \end{subfigure}
  \caption{Linear regression results continued (Panel B).}
\end{sidewaysfigure}
\clearpage
\begin{sidewaysfigure}
\centering
\ContinuedFloat
  \begin{subfigure}[b]{1.03\textheight}
    \centering
    \includegraphics[trim=5 5 5 5,clip,width=\textwidth,height=0.8\textwidth,keepaspectratio]{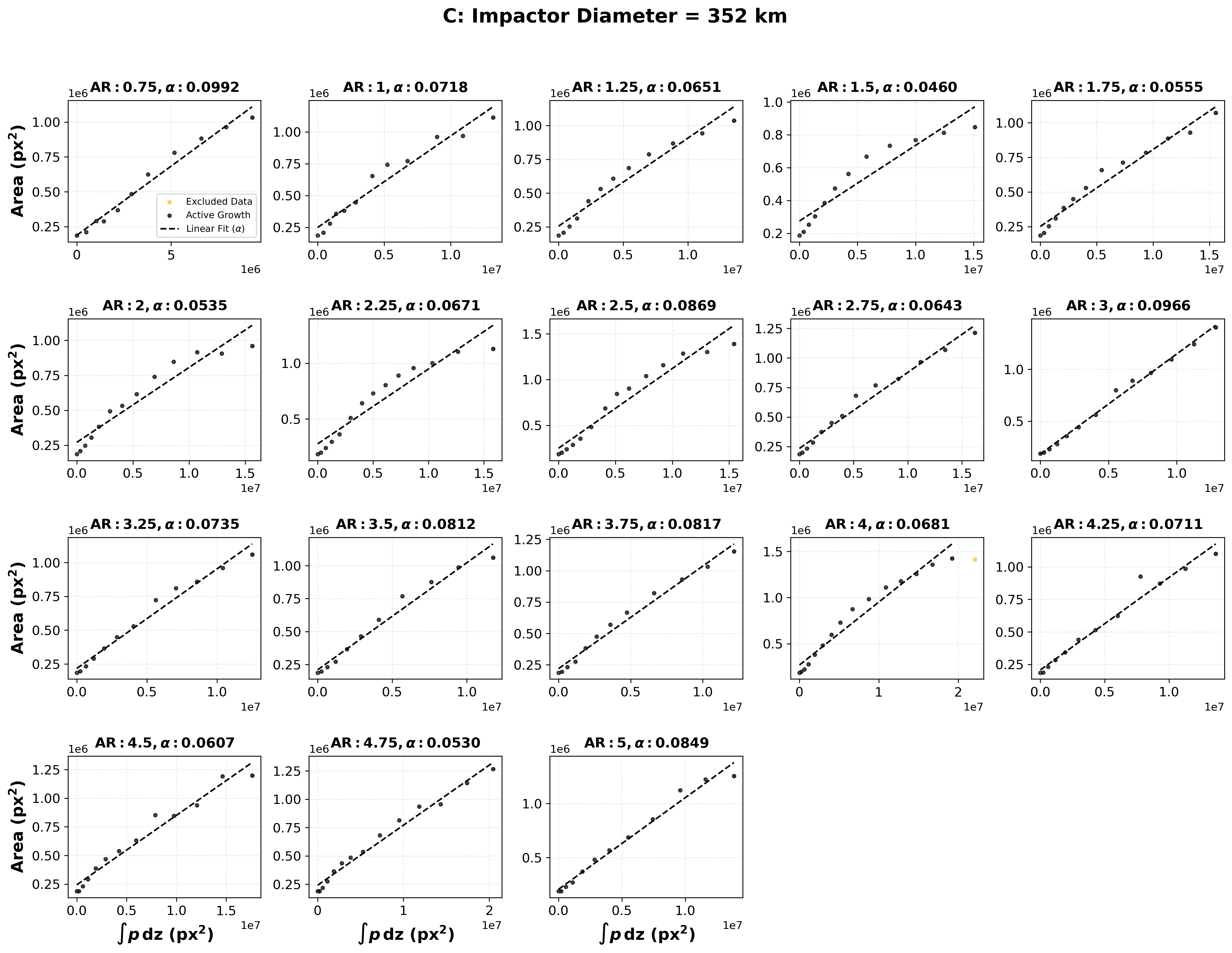}
    \label{subfig:alpha_352}
  \end{subfigure}
  \caption{Linear regression results continued (Panel C).}
\end{sidewaysfigure}
\clearpage

\begin{sidewaysfigure}
\centering
\ContinuedFloat
  \begin{subfigure}[b]{1.03\textheight}
    \centering
    \includegraphics[trim=5 5 5 5,clip,width=\textwidth,height=0.8\textwidth,keepaspectratio]{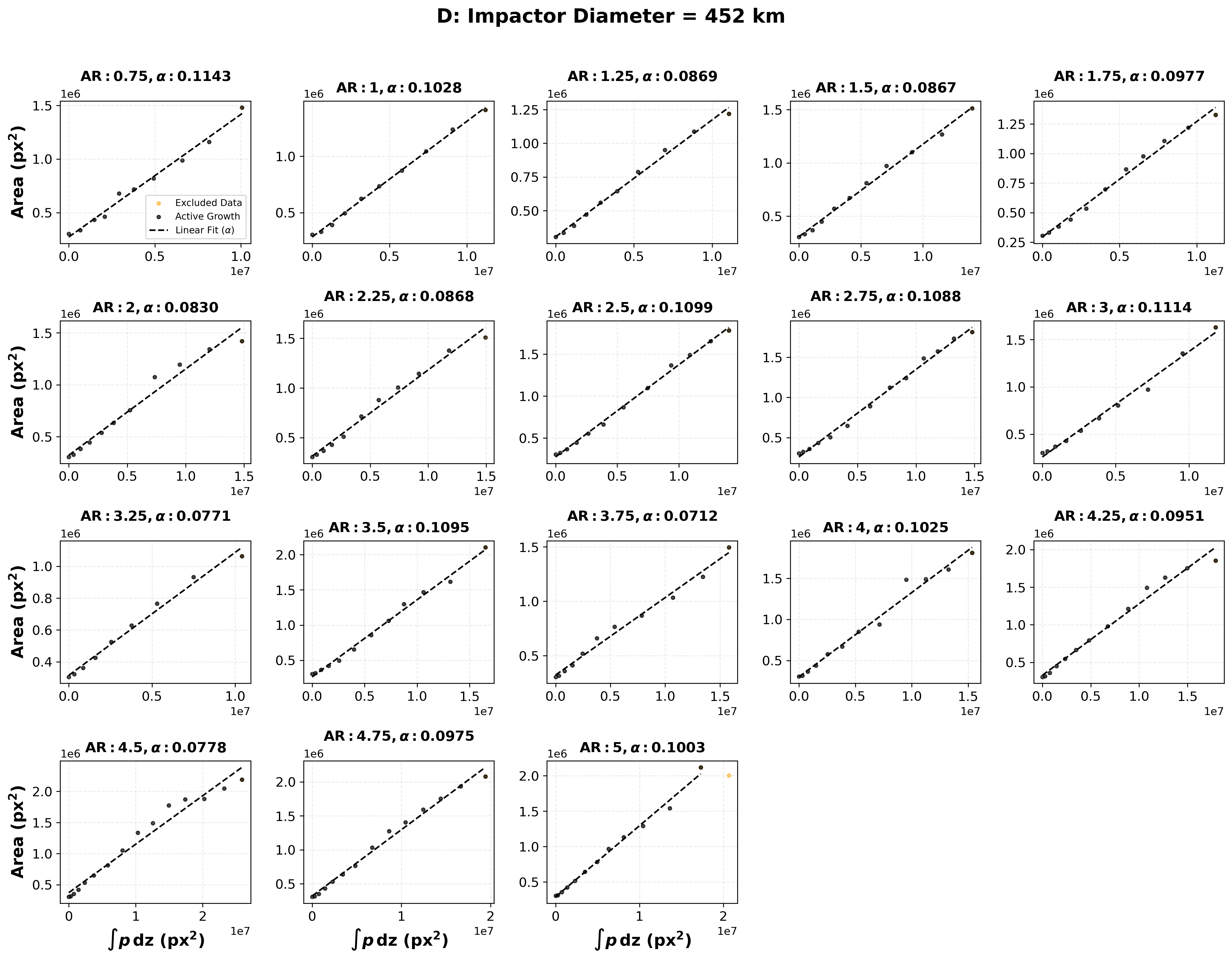}
    \label{subfig:alpha_452}
  \end{subfigure}
  \caption{Linear regression results continued (Panel D).}
\end{sidewaysfigure}
\clearpage

\begin{sidewaysfigure}
\centering
\ContinuedFloat
  \begin{subfigure}[b]{1.03\textheight}
    \centering
    \includegraphics[trim=5 5 5 5,clip,width=\textwidth,height=0.8\textwidth,keepaspectratio]{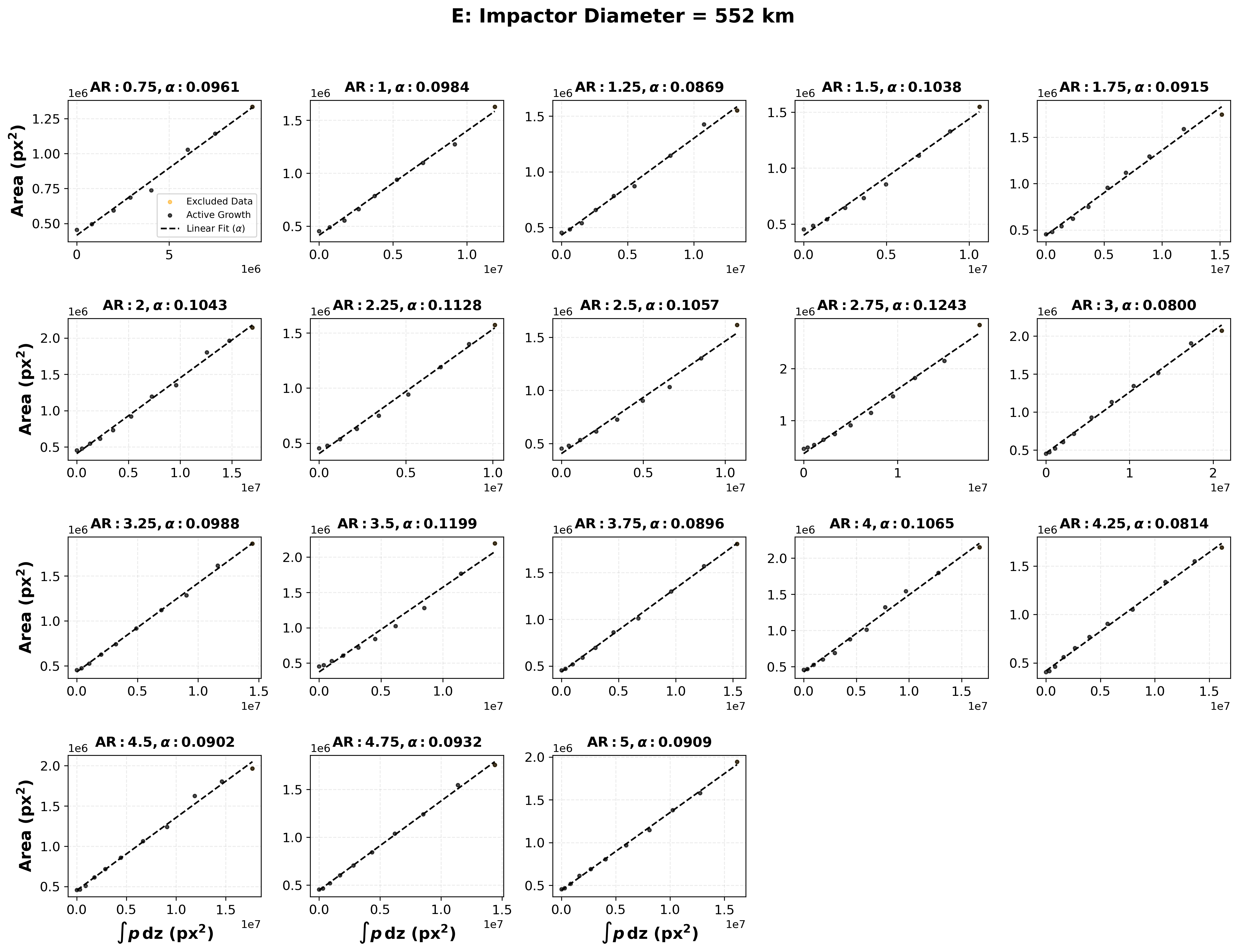}
    \label{subfig:alpha_552}
  \end{subfigure}
  \caption{Linear regression results continued (Panel E).}
\end{sidewaysfigure}

\section*{Instantaneous Mixing Dynamics for $50\;\text{km}$ and $1100\;\text{km}$ Impactors with Aspect Ratios of $1$ and $5$}

\begin{figure}[htbp]
  \centering
  \includegraphics[trim=5 5 5 5,clip,width=1.0\textwidth,keepaspectratio]{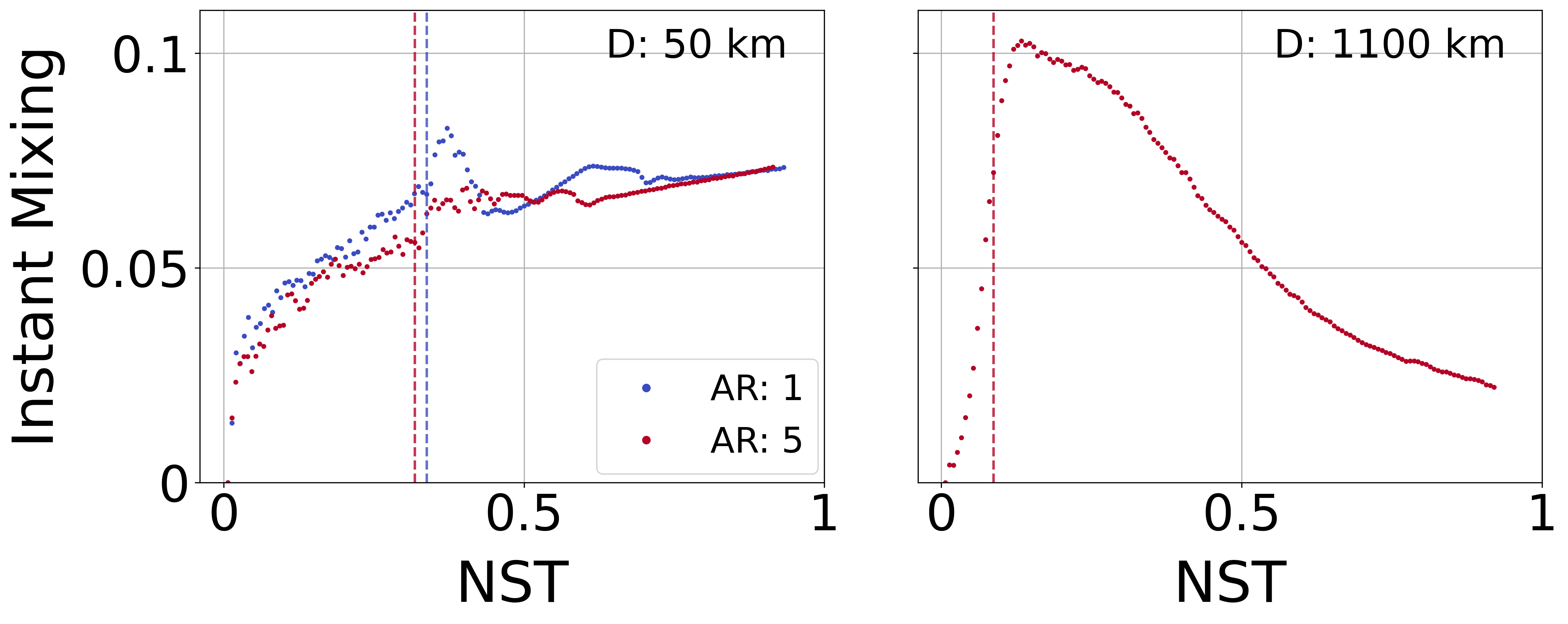}
  \caption{Temporal evolution of an instant mixing for an impactor with a diameter of $50\;\text{km}$ and aspect ratios of $1$ and $5$ (left panel), and an impactor with a diameter of $1100\;\text{km}$ and an aspect ratio of $5$ (right panel). Dashed lines mark the time at which the impactors initially reached the bottom of the MO.}
  \label{fig:shannon_larger}
\end{figure}

Figure \ref{fig:shannon_larger} shows the evolution of instant mixing over the simulation time for impactors smaller and larger than the intermediate size range in this study.

\end{appendices}


\bibliographystyle{cas-model2-names}

\bibliography{cas-refs}



\end{document}